\documentclass[11pt,a4paper]{article}

\usepackage{a4}
\usepackage{amsmath}
\usepackage{amssymb}
\usepackage{axodraw2}
\usepackage{bbm}
\usepackage{color}
\usepackage{graphicx}
\usepackage{jheppub}
\usepackage{latexsym}
\usepackage{mathtools}
\usepackage{pifont}
\usepackage{slashed}

\newcommand{\Neps}{n_\epsilon}
\newcommand{\alphas}{\alpha_s}
\newcommand{\alphae}{\alpha_e}

\newcommand{\QS}[1]{\text{QS}_{[#1]}}
\newcommand{\betaMS}[1]{\bar\beta^s_{#1}}
\newcommand{\betaeMS}[1]{\bar\beta^e_{#1}}

\def\A{{\scshape a}}
\def\V{{\scshape v}}
\def\R{{\scshape r}}
\def\L{{\scshape l}}
\def\CDR{{\scshape cdr}}
\def\DRED{{\scshape dred}}

\def\FDF{{\scshape fdf}}
\def\FDH{{\scshape fdh}}

\def\IREG{{\scshape ireg}}
\def\HV{{\text{\scshape{hv}}}}
\def\HVBM{{\text{\scshape{bm}}}}

\def\SDR{{\scshape sdr}}

\def\GS{{\text{\scshape{gs}}}}
\def\AB{{\text{\scshape{abj}}}}
\def\AC{{\text{\scshape{ac}}}}
\def\G{{\text{\scshape{g}}}}
\def\J{{\text{\scshape{j}}}}
\def\MS{{\overline{\text{\scshape{ms}}}}}
\def\A{{\text{\scshape{a}}}}
\def\dim{{d}}

\allowdisplaybreaks
\begin{document}
\thispagestyle{empty}

\begin{flushright}
PSI-PR-17-16\\
ZU-TH 28/17\\
\end{flushright}
\vspace{3em}
\begin{center}
{\Large\bf $\gamma_5$ in FDH}
\\
\vspace{3em}
{\sc
C.\,Gnendiger$^{a,}$\footnote{e-mail: Christoph.Gnendiger@psi.ch},
A.\,Signer$^{a,b}$
}\\[2em]
{\sl ${}^a$ Paul Scherrer Institut,\\
CH-5232 Villigen PSI, Switzerland \\
\vspace{0.3cm}
${}^b$ Physik-Institut, Universit\"at Z\"urich, \\
Winterthurerstrasse 190,
CH-8057 Z\"urich, Switzerland}
\setcounter{footnote}{0}
\end{center}
\vspace{6ex}

\begin{center}
\begin{minipage}{15.3truecm}
{}We investigate the regularization-scheme dependent treatment of $\gamma_5$
in the framework of dimensional regularization, mainly focusing on the
four-dimensional helicity scheme (\FDH). Evaluating distinctive examples,
we find that for one-loop calculations, the recently proposed four-dimensional
formulation (\FDF) of the \FDH\ scheme constitutes a viable and efficient
alternative compared to more traditional approaches.
In addition, we extend the considerations to the two-loop level and compute
the pseudo-scalar form factors of quarks and gluons in \FDH. We provide the
necessary operator renormalization and discuss at a practical level how the
complexity of intermediate calculational steps can be reduced in an efficient way.
\end{minipage}
\end{center}

\vspace{0.5cm}
\centerline

\newpage
\setcounter{page}{1}

 \noindent\hrulefill
 \tableofcontents
 \noindent\hrulefill

\bigskip
\section{Introduction}
\label{sec:introduction}

The success of quantum-field theoretical predictions over the past decades was
enabled, among other things, by the applicability of dimensional regularization
as the method provides a mathematically consistent tool to handle ultraviolet
(UV) and infrared (IR) divergences in the multi-loop regime.
From the very moment of the introduction of dimensional regularization in
Ref.\,\cite{'tHooft:1972fi}, however, special attention had to be paid to
the treatment of $\gamma_5$ since the object is closely related to concepts
that are only valid in integer dimensions. In a series of publications
\cite{Breitenlohner:1977hr,Jones:1982zf,Korner:1991sx,Kreimer:1993bh,
Larin:1993tq,Schubert:1993wg,Trueman:1995ca,Jegerlehner:2000dz,Tsai:2009hp,
Tsai:2009it,Ferrari:2014jqa,Ferrari:2015mha,Moch:2015usa,Ferrari:2016nea}
that cover a time span of more than 40 years, different approaches have been
developed in order to find consistent rules for the treatment of $\gamma_5$
in the dimensional framework. Irrespective of this effort, in the
overwhelming majority of computations that have been performed so far, the
original $\gamma_5$ definition of Ref.\,\cite{'tHooft:1972fi} has been
used, giving expression to the fact that even today no efficient
alternatives are available that are well suited for all kinds of calculations.

Parallel to the development of $\gamma_5$ schemes, the search for new
efficient calculational methods has focused on finding regularization
prescriptions that reduce the technical complexity at the practical level.
Recently, the current status of the most prominent schemes has been summarized
in Ref.\,\cite{Gnendiger:2017pys}. Among the considered dimensional schemes are
the 't Hooft-Veltman scheme (\HV)
  \cite{'tHooft:1972fi},
conventional dimensional regularization (\CDR)
  \cite{Collins:1984xc},
dimensional reduction (\DRED)
  \cite{Siegel:1979wq},
the four-dimensional helicity scheme (\FDH)
  \cite{BernZviKosower:1992, Bern:2002tk}, and
its recently proposed four-dimensional formulation (\FDF)
  \cite{Fazio:2014xea} at one loop.

In this article, we investigate the treatment of $\gamma_5$ in the aforementioned
dimensional schemes, mainly concentrating on the \FDH\ scheme. As prescriptions
for $\gamma_5$ we consider the original one of 't Hooft/Veltman and an
anticommuting $\gamma_5$. Having the practitioner in mind, we perform distinctive
one- and two-loop calculations and show which of the $\gamma_5$ schemes is the
more efficient alternative for the respective process under consideration.
In order to enable a step-by-step comparison between the different $\gamma_5$
schemes and the different dimensional schemes, the outline of the letter is the
following:
In Sec.\,\ref{sec:g5CDR}, we provide the definitions of $\gamma_5$ in
\CDR/\HV\ and extend them to \FDH/\DRED\ in Sec.\,\ref{sec:g5FDH}.
To illustrate practical consequences of these definitions, we evaluate
characteristic one-loop examples in Secs.\,\ref{sec:oneLoopExample1} and
\ref{sec:oneLoopExample2}, putting emphasis on differences and similarities
of the various approaches.
The extension of these considerations to the two-loop level is discussed in
Sec.\,\ref{sec:psFF} by computing the pseudo-scalar form factors of quarks
and gluons in massless QCD.
The necessary operator renormalization as well as the UV-renormalized
results are provided in Sec.\,\ref{sec:uv}.
\\
\section{Treatment of $\gamma_5$ in dimensional regularization}
\label{sec:g5}

\subsection{CDR and HV}
\label{sec:g5CDR}

One main reason for the recurrent appearance of seeming inconsistencies
related to $\gamma_5$ is the fact that for a consistent formulation
of $d$-dimensional integration, the four-dimensional Minkowski
space $S_{[4]}$ has to be embedded into an \textit{infinite}-dimensional
space $\QS{d}$~\cite{Collins:1984xc},
\begin{align}
 S_{[4]}\subset\QS{d}\,.\text{\footnotemark}
\end{align}
\footnotetext{Following Ref.\,\cite{Gnendiger:2017pys}, we denote the
  (quasi)dimensionality $dim$ of a quantity by a subscript $[dim]$.
  Throughout this article, the modified space-time dimension is always
  defined as $d\!\equiv\!4\!-\!2\epsilon$.}%
Although $\QS{d}$ and the related quantities formally have finite-dimensional
properties, common concepts of $S_{[4]}$ like index counting are no longer
applicable. Regarding $\gamma_5$, this interplay between finite- and
infinite-dimensional aspects has caused quite a lot of confusion in the past
and led to the introduction of different $\gamma_5$ schemes (\GS).

Depending on which \GS\ is chosen, special attention has
to be paid to the evaluation of the Lorentz algebra,
to the breaking of symmetries,
to the treatment of anomalies, and
to the UV renormalization at higher perturbative orders.
According to the different characteristics regarding these points,
it is useful to distinguish the following two classes of \GS:
\begin{itemize}
 \item The first class contains schemes
  where $\gamma_5$ is defined by
  a \textit{construction prescription} like in the original
  definition by 't~Hooft/Veltman \cite{'tHooft:1972fi} and
  Breitenlohner/Maison (\HVBM) \cite{Breitenlohner:1977hr},
  \begin{align}
    \underline{\HVBM}\text{\footnotemark}\!:
    \quad\gamma_{5}^{\HVBM}
    \equiv\frac{i}{4!}\,\big(\,\varepsilon^{\mu\nu\rho\sigma}\,
      \gamma_{\mu}\gamma_{\nu}\gamma_{\rho}\gamma_{\sigma}\big)_{[4]}
    \equiv\frac{i}{4!}\,\varepsilon^{\mu\nu\rho\sigma}_{[4]}\,
    \big(\gamma_{\mu}\gamma_{\nu}\gamma_{\rho}\gamma_{\sigma}\big)_{[\dim]}
    \,.\label{eq:g5hvbm}
  \end{align}
  \footnotetext{In order to distinguish this prescription from other
    aspects of the original \HV\ scheme, we solely use the abbreviation
    \HVBM\ to denote a scheme for the treatment of $\gamma_5$.}
 \item The second class contains schemes where $\gamma_5$ is defined
  \textit{algebraically}, for example as anticommuting (\AC) with (quasi)
  $\dim$-dimensional $\gamma$ matrices~\cite{Korner:1991sx,Kreimer:1993bh},
  \begin{align}
    \underline{\AC}:\quad\{\gamma_{5}^{\AC},\gamma^{\mu}_{[\dim]}\}
    \equiv 0\,.
    \label{eq:g5ac}
  \end{align}
\end{itemize}
In Eq.\,\eqref{eq:g5hvbm}, $\gamma_5^{\HVBM}$ is defined via the totally
antisymmetric Levi-Civita pseudotensor $\varepsilon^{\mu\nu\rho\sigma}$
which is closely related to the concept of index counting in strictly four
dimensions. While the dimensionality of the $\gamma$ matrices is treated
differently in various dimensional schemes, it is mandatory to consider
$\varepsilon^{\mu\nu\rho\sigma}$ as a \textit{strictly} four-dimensional
object. Only in this way it is possible to avoid ambiguous results and
mathematical inconsistencies found before e.\,g.\ in Ref.\,\cite{Siegel:1980qs}.
Usually, the mismatch between the dimensionality of
$\varepsilon^{\mu\nu\rho\sigma}$ and other algebraic objects is circumvented
by workarounds whose ranges of validity are often not obvious, at least not
at first sight. More details regarding this issue will be given in
Sec.\,\ref{sec:psFFcommon}.

A direct consequence of Eq.\,\eqref{eq:g5hvbm} is that all
(anti)commutation relations of $\gamma_5^{\HVBM}$ are implicitly
part of the definition and therefore fixed, e.\,g.\
\begin{subequations}
\label{eq:g5hvRules}
\begin{align}
 \big\{\gamma_{5}^{\HVBM},\,\gamma^{\mu}_{[4]}\big\}=0\,,\qquad
 \big[\,\gamma_{5}^{\HVBM},\,\gamma^{\mu}_{[d-4]}\,\big]=0\,, \,
 \label{eq:g5hvComm}
\end{align}
and therefore \cite{Breitenlohner:1977hr}
\begin{align}
 \big\{\gamma_{5}^{\HVBM},\,\gamma^{\mu}_{[d]}\big\}
  =2\,\gamma_{[d-4]}^{\mu}\,\gamma_{5}^{\HVBM} \,.
 \label{eq:g5hvneq}
\end{align}
\end{subequations}
It is clear that Eqs.\,\eqref{eq:g5ac} and \eqref{eq:g5hvneq} yield
different results for $d\!\neq\!4$, at least at intermediate steps
of the calculation. In the UV renormalized (and IR subtracted) theory,
however, different consistent approaches have to yield the same results
for physical observables.\\

\subsection{FDH and DRED}
\label{sec:g5FDH}

So far, the algebraic behavior of $\gamma_5$ has been considered in the
quasi $d$-dimensional space $\QS{d}$ which is the natural domain of
\CDR\ and of $d$-dimensional integration.
In Ref.\,\cite{Stockinger:2005gx}, it is shown that in order to
consistently formulate \FDH\ and \DRED, this space has to be enlarged
to $\QS{d_s}$ via a direct (orthogonal) sum with the so-called
'evanescent' space $\QS{\Neps}$,
\begin{align}
 \QS{d_s}\equiv\QS{d}\oplus \QS{\Neps} \, .
 \phantom{\bigg|}
 \label{eq:vsDecomp}
\end{align}
Although $d_s$ is usually taken to be $4$ in \FDH\ and \DRED, it is clear
that $\QS{d_s}$ is an infinite-dimensional space with finite-dimensional
algebraic properties.%
\footnote{For more comments on the definition and the structure of the
  vector spaces in Eq.\,\eqref{eq:vsDecomp} we refer to~\cite{Stockinger:2005gx,
  Gnendiger:2014nxa, Broggio:2015dga} and references therein. Here it should
  only be mentioned that setting $\dim_s\!=\!4$ results in $\Neps\!=\!2\epsilon$.}

According to the structure of the vector spaces in Eq.\,\eqref{eq:vsDecomp},
quasi $d_s$-dimensional metric tensors and $\gamma$ matrices can be split as
$g_{[d_s]}^{\mu\nu}=g_{[d]}^{\mu\nu}+g_{[\Neps]}^{\mu\nu}$ and
$\gamma_{[d_s]}^{\mu}=\gamma_{[\dim]}^{\mu}+\gamma_{[\Neps]}^{\mu}$,
resulting in
\begin{subequations}
\label{eq:FDHalg}
\begin{align}
  \big(g_{[dim]}\big)^{\mu}_{\phantom{\mu}\mu}
    &\ =\ dim\,,&
  \big(g_{[\dim]}\,g_{[\Neps]}\big)^{\mu}_{\phantom{\mu}\nu}
    &\ =\ 0\,,&
  \phantom{\Big|}
\\*
  \{\gamma_{[dim]}^{\mu},\,\gamma_{[dim]}^{\nu\phantom{\mu}}\}
    &\ =\ 2\,g_{[dim]}^{\mu\nu}\,,&
  \{\gamma_{[\dim]}^{\mu},\,\gamma_{[\Neps]}^{\nu\phantom{\mu}}\}
    &\ =\ 0\,,&
  \phantom{\Big|}
  \label{eq:algRel}
\end{align}
\end{subequations}
with $dim\in\{4,\,\dim,\,d_s,\Neps\}$.
\newpage

\noindent
As mentioned before, the (anti)commutation relations of $\gamma_{5}^{\HVBM}$
are fixed by Eq.\,\eqref{eq:g5hvbm}, e.\,g.\ 
\begin{subequations}
\begin{align}
 \underline{\HVBM}:
 \qquad\big\{\gamma_{5}^{\HVBM},\,\gamma^{\mu}_{[d]}\big\}
  &= 2\,\gamma_{[d-4]}^{\mu}\,\gamma_{5}^{\HVBM}\,,&
  \big[\,\gamma_{5}^{\HVBM},\,\gamma^{\mu}_{[\Neps]}\,\big]
  &= 0\,.&
 \phantom{\bigg|}
 \label{eq:g5hvneq2}
\end{align}
Due to the even number of $\gamma$ matrices in Eq.\,\eqref{eq:g5hvbm},
$\gamma_{5}^{\HVBM}$ \textit{commutes} with the evanescent degrees of
freedom in \FDH\ and \DRED. Moreover, from Eq.\,\eqref{eq:g5hvneq2} it
directly follows that the structure of the (anti)commutation relation
in $\dim$ and $\dim_s$ dimensions is the same,
\begin{align}
 \big\{\gamma_{5}^{\HVBM},\,\gamma^{\mu}_{[d_s]}\big\}
  &= 2\,\gamma_{[d_s-4]}^{\mu}\,\gamma_{5}^{\HVBM}\,.
 \phantom{\bigg|}
 \label{eq:g5hvneq3}
\end{align}
\end{subequations}
As a consequence, in practical calculations it is possible
to either use a quasi $d_s$-dimensional Lorentz algebra or to
explicitly perform the split of Eq.\,\eqref{eq:vsDecomp}.

In contrast, the (anti)commutation relations of $\gamma_5^{\AC}$
are not fixed a priori but have to be part of the definition.
We therefore \textit{define}
\begin{subequations}
\label{eq:g5acComm}
\begin{align}
 \underline{\AC}:\qquad
  \big\{\gamma_{5}^{\AC},\,\gamma^{\mu}_{[d]}\big\}&\equiv0\,,&
  \big\{\gamma_{5}^{\AC},\,\gamma^{\mu}_{[\Neps]}\big\}&\equiv0\,,&
  \label{eq:g5hvneq2a}
\end{align}
resulting in
\begin{align}
  \big\{\gamma_{5}^{\AC},\,\gamma^{\mu}_{[d_s]}\big\}&=0\,.
  \label{eq:g5hvneq2b}
\end{align}
\end{subequations}
At first sight, it might seem appropriate to use a \textit{commutator}
in the right definition of Eq.\,\eqref{eq:g5hvneq2a}, in a similar way
as in Eq.\,\eqref{eq:g5hvneq2}.
In general, however, calculations in \FDH\ and \DRED\ are significantly
facilitated if one uses a quasi $d_s$-dimensional algebra instead of
performing the split in Eq.\,\eqref{eq:vsDecomp}. This option is guaranteed by
Eqs.\,\eqref{eq:g5acComm} since the algebra in $\dim$ and $\dim_s$ dimensions is
the same. Moreover, in Secs.\,\ref{sec:oneLoopExample1} and \ref{sec:acG5uv} it
will be shown that exclusively using anticommutators in Eqs.\,\eqref{eq:g5acComm}
results in a much simpler UV renormalization. It is also a convenient choice
regarding the non-breaking of supersymmetry~\cite{Stockinger:2005gx}.

To illustrate the implications of the different schemes for $\gamma_5$,
we consider the following simple one-loop examples in the \FDH\ scheme:
the correlator $\gamma^{\mu}\gamma_5\to e^{+}e^{-}$ and the (anomalous)
AVV triangle.
Each of the examples is evaluated by using $\gamma_5^{\HVBM}$ and
$\gamma_5^{\AC}$ as defined in Eqs.\,\eqref{eq:g5hvbm} and \eqref{eq:g5ac},
respectively. In addition we apply \FDF, a recently proposed genuine
four-dimensional formulation of the \FDH\ algebra at the one-loop level.
In the analytical results, the fermion mass is denoted by $m$ and $p_1, p_2$
are the (outgoing) momenta of the external fermions/gauge fields. For
simplicity we consider QED and set $e\!=\!1$ for the gauge coupling.\\

\subsection{One-loop example 1:
correlator $\gamma^{\mu}\gamma_5\to e^{+}e^{-}$}
\label{sec:oneLoopExample1}

\subsubsection*{FDH and $\gamma_{5}^{\text{BM}}$}

\begin{figure}[t]
\begin{center}
\scalebox{.75}{
\begin{picture}(135,90)(0,10)
\end{picture}
\quad
\begin{picture}(135,90)(0,10)
 \DashLine(0,45)(30,45){2}
 \Line[arrow](30,45)(100,80)
 \Line[arrow](100,80)(120,90)
 \Line[arrow](120,0)(100,10)
 \Line[arrow](100,10)(30,45)
 \Photon(100,10)(100,80){6}{6}
 \Vertex(30,45){2}
 \Vertex(100,80){2}
 \Vertex(100,10){2}
 \Text(20,60)[c]{\scalebox{1.33}{$\gamma^{\mu}\gamma_{5}$}}
\end{picture}
\qquad\qquad\quad
\begin{picture}(135,90)(0,10)
 \DashLine(0,45)(30,45){2}
 \Line[arrow](30,45)(100,80)
 \Line[arrow](100,80)(120,90)
 \Line[arrow](120,0)(100,10)
 \Line[arrow](100,10)(30,45)
 \DashLine(100,10)(100,80){6}
 \Vertex(30,45){2}
 \Vertex(100,80){2}
 \Vertex(100,10){2}
 \Text(20,60)[c]{\scalebox{1.33}{$\gamma^{\mu}\gamma_{5}$}}
\end{picture}
\quad
\begin{picture}(135,90)(0,10)
\end{picture}
}
\end{center}
\caption{\label{fig:fig2}
One-loop contributions to the correlator
$\gamma^{\mu}\gamma_5\to e^{+}e^{-}$.
The diagrams contain a gauge field (left) and an associated \FDF-scalar
(right). The latter diagram is only present in \FDF.
}
\end{figure}
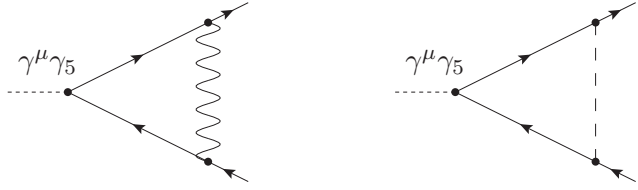

The application of $\gamma_{5}^{\text{\HVBM}}$ in a $\dim$-dimensional framework
with $\dim\!\neq\!4$ results in different algebraic properties compared to the
unregularized theory which can be easily seen from Eqs.\,\eqref{eq:g5hvRules}.
The ($\dim$-dimensional) axial-vector operator is therefore usually
symmetrized 'by hand' and written as~\cite{Larin:1993tq}
\begin{align}
 \gamma^{\mu}_{[4]}\,\gamma_{5}^{}\to\frac{1}{2}\,\big(
   \gamma^{\mu}_{[\dim]}\,\gamma_{5}^{\HVBM}
   \!-\!\gamma_{5}^{\HVBM}\,\gamma^{\mu}_{[\dim]}\big)\, .
 \label{eq:g5symm}
\end{align}
Using this relation together with Eqs.\,\eqref{eq:g5hvbm} and \eqref{eq:FDHalg},
and multiplying with $q_{\mu}\!\equiv\!(p_{1}\!+\!p_{2})_{\mu}$ then yields for
the left diagram in Fig.\,\ref{fig:fig2}%
\footnote{In this example, Lorentz indices related to vector fields
  are treated in $\dim_s\!=\!\dim\!+\!\Neps$ dimensions, see also
  Eq.\,\eqref{eq:vsDecomp}. The case $\dim_s\!=\!4$ (and therefore
  $\Neps\!=\!2\epsilon$) then corresponds to \FDH\ and \DRED,
  whereas results in \CDR\ and \HV\ are obtained for $\Neps\!=\!0$.
  Here and in the following, the irrelevant dimension of the external
  momenta is set to $\dim$ and terms of $\mathcal{O}(\epsilon^0\Neps)$
  are omitted since they vanish after setting $\Neps\!=\!2\epsilon$ and
  taking the subsequent limit $\epsilon\!\to\!0$. The regularization scale
  is fixed via $\mu_0\!\equiv\!m$. Note further, that the $\varepsilon$
  pseudotensor is considered outside dimensional regularization and
  treated in strictly four dimensions.}
\begin{align}
q_{\mu}&\,T^{\mu}\big|_{\text{bare}}\to
\notag\\&
 \frac{\varepsilon^{\mu\nu\rho\sigma}_{[4]}}{2\times4!}
 \!\int\!\frac{d^{\dim}k}{(2\pi)^{\dim}}\frac{
 \gamma^{\alpha}_{[\dim_s]}\,
 \big[
  \big(
    \slashed{k}\!+\!\slashed{p}_{1}\!+\!m\big)\,
   \big(
  \slashed{q}\,
  \gamma_{\mu}
  \gamma_{\nu}
  \gamma_{\rho}
  \gamma_{\sigma}
  \!-\!
  \gamma_{\mu}
  \gamma_{\nu}
  \gamma_{\rho}
  \gamma_{\sigma}\,
  \slashed{q}
 \big)
  \big(
    \slashed{k}\!-\!\slashed{p}_{2}\!+m\!\big)
  \big]_{[\dim]}
  \gamma^{\beta}_{[\dim_s]}\,
  \big(g_{\alpha\beta}^{\phantom{\beta}}\big)_{[\dim_s]}
  }{
  \big[(k\!+\!p_{1})^2_{[\dim]}\!-\!m^2\big]
  \big[(k\!-\!p_{2})^2_{[\dim]}\!-\!m^2\big]\,
  k^2_{[\dim]}}
  \notag\\*&\quad\,
 =\frac{1}{(4\pi)^2}\,
  \bigg[\,
    \frac{1\!-\!\frac{\Neps}{2}}{\epsilon}
    +\frac{9}{2}
    +\mathcal{O}(\epsilon)
    +\mathcal{O}(m^2)\,\bigg]
 \,\slashed{q}_{[\dim]}\gamma_5^{\HVBM}\,.
 \label{eq:ex2fdhbms}
\end{align}
The (on-shell) renormalization of the external fermion fields as well as the
prediction for the structure of the IR divergences in the \FDH\ scheme are
given in Ref.\,\cite{Gnendiger:2016cpg},
\begin{subequations}
\begin{align}
 \delta \bar{Z}^{(1)}_2(\Neps)
 &=\frac{1}{(4\pi^2)}
  \Big[\frac{-3-\frac{\Neps}{2}}{\epsilon}-4
  +\mathcal{O}(\epsilon)+\mathcal{O}(m^2)\Big]\,,
  \label{eq:Z2fdh}
 \\
 \bar{\mathbf{Z}}^{(1)}_{\text{IR}}
 &=\frac{1}{(4\pi^2)}\Big[\!-\!\frac{2}{\epsilon}+\mathcal{O}(m^2)\Big]\,.
\end{align}
\end{subequations}
Subtracting the IR divergence, it follows that field renormalization is not
sufficient to obtain the correct result since the (scheme-dependent) UV
divergence does not cancel. The general reason is that symmetries of the
unregularized theory like chiral and Lorentz invariance are broken explicitly
if $\gamma_5^{\HVBM}$ is used in a $\dim$-dimensional framework.%
\footnote{In the original reference of 't~Hooft/Veltman~\cite{'tHooft:1972fi},
  for example, it is shown how the use of Eq.\,\eqref{eq:g5hvneq}
  leads to a breaking of Ward identities. See also Ref.\,\cite{Larin:1993tq}
  for a pedagogical review.}
As a consequence, initial symmetries have to be restored by means of
additional counterterms. In Sec.\,\ref{sec:bmG5uv}, it will be shown that for
the one-loop example at hand, this renormalization reads
\begin{align}
 \delta\bar{Z}^{\HVBM,(1)}(\Neps)
 =\delta\bar{Z}_{\MS}^{\HVBM,(1)}(\Neps)+\delta Z_{5}^{(1)}
 =\frac{1}{(4\pi)^2}\Big[\,
  \frac{\Neps}{\epsilon}-4 
 \,\Big]
 \,.
 \label{eq:ex2fdhbmsd}
\end{align}
It is given by a pure $\MS$ pole term $\delta\bar{Z}^{\HVBM}_{\MS}$ which
is finite after setting $\Neps\!=\!2\epsilon$ and by a regularization-scheme
independent constant $\delta Z_{5}$. In \CDR\ ($\Neps\!=\!0$), the latter is
usually determined through relations that are valid in strictly
four-dimensional schemes like the Pauli-Villars setup,
see e.\,g.\ Ref.\,\cite{Larin:1993tq}.
In Sec.\,\ref{sec:uv} we present an alternative approach that is based
on a comparison between results obtained with $\gamma_5^{\HVBM}$ and
$\gamma_5^{\AC}$.

Combining Eqs.\,\eqref{eq:ex2fdhbms}--\eqref{eq:ex2fdhbmsd}
and taking the subsequent limit $\dim\!\to\!4$, we obtain for the
UV-renormalized and IR-subtracted correlator
\begin{align}
 q_{\mu}\,T^{\mu}&=
 \frac{1}{(4\pi)^2}\bigg[\!-\!\frac{7}{2}+\mathcal{O}(m^2)\bigg]\,
 \slashed{q}_{[4]}\gamma_5\,.
 \label{eq:ex2fdhFinal}
\end{align}
Since all evanescent terms $\sim\Neps$ drop out through UV
renormalization, this final result does not depend on the applied
dimensional scheme.

\subsubsection*{FDH and $\gamma_{5}^{\text{AC}}$}

For the case of an anticommuting $\gamma_5^{\AC}$ we write
the \FDH\ one-loop amplitude as
\begin{align}
q_{\mu}\,T^{\mu}\big|_{\text{bare}}&\to
 -i\!\int\!\frac{d^{\dim}k}{(2\pi)^{\dim}}\frac{
 \gamma^{\alpha}_{[\dim_s]}\,
 \big[
  \big(
    \slashed{k}\!+\!\slashed{p}_{1}\!+\!m\big)\,
  \slashed{q}
  \gamma_5^{\AC}
  \big(
    \slashed{k}\!-\!\slashed{p}_{2}\!+\!m\big)
  \big]_{[\dim]}
  \gamma^{\beta}_{[\dim_s]}\,
  \big(g_{\alpha\beta}^{\phantom{\beta}}\big)_{[\dim_s]}
  }{
  \big[(k\!+\!p_{1})^2_{[\dim]}\!-\!m^2\big]
  \big[(k\!-\!p_{2})^2_{[\dim]}\!-\!m^2\big]\,
  k^2_{[\dim]}}
\notag\\*&\quad\,
 =\frac{1}{(4\pi)^2}\,
 \bigg[\,
  \frac{1\!+\!\frac{\Neps}{2}}{\epsilon}
  +\frac{1}{2}+\mathcal{O}(\epsilon)+\mathcal{O}(m^2)
  \,\bigg]\,
 \slashed{q}_{[\dim]}\gamma_5^{\AC}\,.
 \label{eq:ex2fdhac}
\end{align}
The result has been obtained by (anti)commuting $\gamma_5^{\AC}$ to the
right and evaluating the remaining algebra by means of Eqs.\,\eqref{eq:FDHalg}.
Due to the absence of an explicit symmetrization and the reduced number
of $\gamma$ matrices in the numerator, the evaluation of the algebra is much
simpler compared to Eq.\,\eqref{eq:ex2fdhbms}. Moreover, the consequent use of
an anticommutator in Eqs.\,\eqref{eq:g5acComm} leads to a sign change of the
$\Neps$ term. Applying the field renormalization of Eq.\,\eqref{eq:Z2fdh}
and subtracting the IR divergence we then directly recover the result in
Eq.\,\eqref{eq:ex2fdhFinal}. In contrast to $\gamma_5^{\HVBM}$ therefore
no symmetry-restoring counterterms are needed to get the correct result.

\subsubsection*{Algebra in genuine four dimensions -- FDF}

\FDF\ is a novel regularization approach
that was introduced to reproduce \FDH\ results at the one-loop level~%
\cite{Fazio:2014xea}. Starting from unregularized analytical expressions,
loop momenta in \FDF\ are shifted as $\slashed{k}_{[4]}\!\to\!
\slashed{k}_{[\dim]}\!\equiv\!\slashed{k}_{[4]}\!+\!i\,\mu\,\gamma_5$
\textit{before} any other algebraic manipulation is performed.
The scale $\mu$ corresponds to the $(\dim\!-\!4)$-dimensional components
of the loop momentum and serves as a regulator for the in general divergent
quasi $\dim$-dimensional loop integrals. By definition, odd powers of
$\mu$ are set to zero, resulting in the useful relation
\begin{align}
 \slashed{k}_{[\dim]}\slashed{k}_{[\dim]}=
 k_{[\dim]}^2=k_{[4]}^2-\mu^2\,.
 \label{eq:FDFshift}
\end{align}
One main advantage of the \FDF\ approach is that the Lorentz algebra
is realized in strictly four dimensions; Eqs.\,\eqref{eq:g5hvbm} and
\eqref{eq:g5ac} are therefore equivalent, i.\,e.\
$\gamma_5^{\HVBM}\!=\!\gamma_5^{\AC}\!\equiv\!\gamma_5$.
Applying this setup, the analytical expression for the left diagram in
Fig.\,\ref{fig:fig2} reads%
\footnote{Using Feynman gauge, the right diagram including a so-called
  \FDF-scalar vanishes according to the rules of \FDF; in other gauges,
  both diagrams in Fig.\,\ref{fig:fig2} contribute. In the latter case,
  the diagrams sum up to the same (gauge-independent) result
  as given in Eq.\,\eqref{eq:ex2fdf}. For more details regarding gauge
  dependence in \FDF\ we refer to Ref.\,\cite{Gnendiger:2017pys}.}
\begin{align}
 q_{\mu}\,T^{\mu}\big|_{\text{bare}}&\to -i\!
 \int\!\frac{d^{\dim}k}{(2\pi)^{\dim}}\frac{
 \big[
 \gamma^{\alpha}\,
  \big(
    \slashed{k}
    \!+\!i\,\mu\,\gamma_5
    \!+\!\slashed{p}_{1}
    \!+\!m\big)\,
  \slashed{q}\gamma_5
  \big(
    \slashed{k}
    \!+\!i\,\mu\,\gamma_5
    \!-\!\slashed{p}_{2}
    \!+\!m\big)\,
  \gamma^{\beta}\,
  g_{\alpha\beta}
  \big]_{[4]}
  }{
  \big[(k\!+\!p_{1})^2_{[\dim]}\!-\!m^2\big]
  \big[(k\!-\!p_{2})^2_{[\dim]}\!-\!m^2\big]\,
  k^2_{[\dim]}
  }
  \notag\\*&\quad\,
 =\frac{1}{(4\pi)^2}\,
 \bigg[\,\frac{1}{\epsilon}+\frac{7}{2}
 +\mathcal{O}(\epsilon)
 +\mathcal{O}(m^2)\,\bigg]\,
 \slashed{q}_{[4]}\gamma_5^{}
 \,.
 \label{eq:ex2fdf}
\end{align}
For the evaluation of the algebra we used Eq.\,\eqref{eq:FDFshift} to
cancel against the denominator, resulting in the $\mu^2$-dependent
'extra integral'\,\cite{Gnendiger:2017pys}
\begin{align}
 I_{3}^{\dim}(\mu^2)=
 \int\!\frac{d^{\dim}k}{(2\pi)^{\dim}}
  \frac{\mu^2}{
  \big[(k\!+\!p_{1})^2_{[\dim]}\!-\!m^2\big]
  \big[(k\!-\!p_{2})^2_{[\dim]}\!-\!m^2\big]\,
  k^2_{[\dim]}
  }
 =\frac{i}{(4\pi)^2}\bigg[
  \frac{1}{2}\!+\!\frac{3}{2}\epsilon\!+\!\mathcal{O}(\epsilon^2)
  \bigg]+\mathcal{O}(m^2)\,.
 \label{eq:muInt}
\end{align}
Although only strictly four-dimensional quantities and an anticommuting
$\gamma_5^{}$ have been used to obtain the result in Eq.\,\eqref{eq:ex2fdf},
the $\gamma_5^{\HVBM}$ result in  Eq.\,\eqref{eq:ex2fdhbms} for
$\Neps\!=\!2\epsilon$ is recovered. The conceptual reason is that within
\FDF, similar relations as in Eq.\,\eqref{eq:g5hvneq} hold, e.\,g.%
\footnote{This relation follows from
  $ \gamma_5\,\slashed{k}_{[\dim]}
 \!=\!\gamma_5\big(\slashed{k}_{[4]}\!+\!i\,\mu\,\gamma_5\big)
 \!=\!\big(\!-\!\slashed{k}_{[4]}\!+\!i\,\mu\,\gamma_5\big)\,\gamma_5
 \!=\!-\slashed{k}_{[\dim]}\,\gamma_5\!+\!2\,i\,\mu$.
 It is important to notice that in practical computations, relations like
 in Eq.\,\eqref{eq:g5FDF} are not used explicitly since quasi $\dim$-dimensional
 quantities are in \FDF\ split into a strictly four-dimensional and a
 $\mu$-dependent part. The $\gamma_5$ matrix is therefore effectively an
 anticommuting one.
 }
\begin{align}
  \underline{\text{\FDF}}:\qquad
 \{\gamma_5,\,\slashed{k}_{[\dim]}\}=2\,i\,\mu\,.
 \label{eq:g5FDF}
\end{align}
To obtain a physical result that is compatible with the symmetries of the
underlying theory we therefore have to add the same counterterms as for
the case of $\gamma_5^{\HVBM}$. Compared to Eq.\,\eqref{eq:ex2fdhbms},
however, the evaluation of the analytical expressions is significantly
simplified.\\

\subsection{One-loop example 2: AVV triangle}
\label{sec:oneLoopExample2}

\begin{figure}[t]
\begin{center}
\scalebox{.75}{
\begin{picture}(135,90)(0,10)
\DashLine(0,45)(30,45){2}
\Line[arrow](30,45)(100,10)
\Line[arrow](100,10)(100,80)
\Line[arrow](100,80)(30,45)
\Photon(100,80)(120,90){4}{2}
\Photon(100,10)(120,0){4}{2}
\Vertex(30,45){2}
\Vertex(100,80){2}
\Vertex(100,10){2}
\Text(20,60)[c]{\scalebox{1.33}{$\gamma^{\mu}\gamma_{5}$}}
\Text(92,90)[c]{\scalebox{1.33}{$\gamma^{\alpha}$}}
\Text(92,-2)[c]{\scalebox{1.33}{$\gamma^{\beta}$}}
\end{picture}
\qquad\quad
\begin{picture}(135,90)(0,10)
\DashLine(0,45)(30,45){2}
\Line[arrow](100,10)(30,45)
\Line[arrow](100,80)(100,10)
\Line[arrow](30,45)(100,80)
\Photon(100,80)(120,90){4}{2}
\Photon(100,10)(120,0){4}{2}
\Vertex(30,45){2}
\Vertex(100,80){2}
\Vertex(100,10){2}
\Text(20,60)[c]{\scalebox{1.33}{$\gamma^{\mu}\gamma_{5}$}}
\Text(92,90)[c]{\scalebox{1.33}{$\gamma^{\alpha}$}}
\Text(92,-2)[c]{\scalebox{1.33}{$\gamma^{\beta}$}}
\end{picture}
}
\end{center}
\caption{\label{fig:fig1}
One-loop contributions to the (anomalous) AVV correlator
$T^{\mu\alpha\beta}_{\text{\A\V\V}}$ including one axial-vector
and two vector vertices.
}
\end{figure}
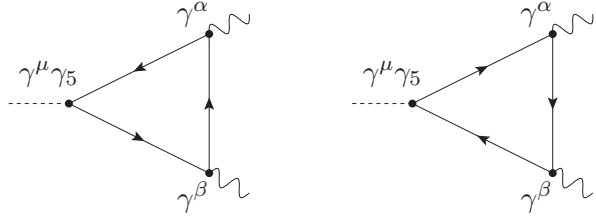

As a second example we consider the AVV triangles in Fig.\,\ref{fig:fig1}
for the case of massless fermions. In the present case of an NLO fermion loop,
the only difference between the dimensional schemes \CDR, \HV, \FDH, and \DRED\
is the dimensionality of the external gauge-field momenta. Since the
final result of the amplitude is finite, as will be shown below, the
limit $\dim\!\to\!4$ can be taken without any UV renormalization. After having
taken the physical limit, the virtual one-loop amplitudes are therefore the same
in all these dimensional schemes.

\subsubsection*{FDH and $\gamma_{5}^{\text{BM}}$}

Applying the same setup as in the previous example
we obtain in \CDR
\begin{align}
 q_\mu & \,T^{\mu\alpha\beta}_{\text{\A\V\V}}
 \to
 \notag\\*&
 \frac{i\,\varepsilon^{\mu\nu\rho\sigma}_{[4]}}{2\times4!}\!
 \int\!\frac{d^{\dim}k}{(2\pi)^{\dim}}\frac{
 \text{Tr}\big[
 \big(
  \slashed{q}\,
  \gamma_{\mu}
  \gamma_{\nu}
  \gamma_{\rho}
  \gamma_{\sigma}
  \!-\!
  \gamma_{\mu}
  \gamma_{\nu}
  \gamma_{\rho}
  \gamma_{\sigma}\,
  \slashed{q}
 \big)
 \big(\slashed{k}\!+\!\slashed{p}_{1}\big)\,
 \gamma^{\alpha}\,
 \slashed{k}\,
 \gamma^{\beta}\,
 \big(\slashed{k}\!-\!\slashed{p}_{2}\big)
 \big]_{[\dim]}
  }{
  (k\!+\!p_{1})^2_{[\dim]}\,
   k^2_{[\dim]}\,
  (k\!-\!p_{2})^2_{[\dim]}
  }
  \!+\!\bigg(\begin{aligned}
    p_1&\!\leftrightarrow\!p_2\\
    \alpha&\!\leftrightarrow\!\beta
   \end{aligned}\bigg)
  \notag\\*&\quad\,
  =-\frac{1}{2\pi^2}\,
    \varepsilon^{\alpha\beta\mu\nu}_{[4]}
    \Big\{p_{1,\mu}\,p_{2,\nu}\Big\}_{[\dim]}
    \Big[\,1\!+\!3\epsilon\!+\!\mathcal{O}(\epsilon^2)\Big]\,,
  \label{eq:qTCDRBM}
\end{align}
where, as before, the $\varepsilon$ pseudotensor is considered outside
dimensional regularization throughout the calculation. Taking the limit
$\epsilon\!\to\!0$, the result in Eq.\,\eqref{eq:qTCDRBM} coincides with the
well-known (anomalous) axial Ward identity (AWI) given e.\,g.\ in
Refs.\,\cite{Adler:1969gk,Bell:1969ts,Adler:1969er}.

\subsubsection*{FDH and $\gamma_{5}^{\text{AC}}$}

One important characteristic related to the treatment of
$\gamma_{5}^{\text{\AC}}$ in dimensional schemes is that traces including
odd numbers of $\gamma_5^{\AC}$ either vanish or are not cyclic anymore.
Demanding, for example, cyclicity of traces including $\gamma_5^{\AC}$
leads to relations like~\cite{Jegerlehner:2000dz}
\begin{align}
 (\dim\!-\!4)\ \text{Tr}\big[
  \gamma^{\mu}
  \gamma^{\nu}
  \gamma^{\rho}
  \gamma^{\sigma}\,
  \gamma_5^{\AC}\big]_{[\dim]}=0\,.
 \label{eq:actrace}
\end{align}
For $d\!\neq\!4$, this equation can only be fulfilled for a vanishing
trace. Since similar relations hold for other numbers of $\gamma$
matrices in the trace we get
\begin{align}
 &q_\mu\,T^{\mu\alpha\beta}_{\text{\A\V\V}}
 = 0
\end{align}
and gauge invariance is broken explicitly. Different solutions have been
proposed e.\,g.\ in Refs.\,\cite{Korner:1991sx, Kreimer:1993bh} and
\cite{Ferrari:2014jqa, Ferrari:2015mha, Ferrari:2016nea} by modifying 
the trace operation in such a way that the result in Eq.\,\eqref{eq:qTCDRBM}
is recovered. As mentioned before, these modified traces are not cyclic which
leads to significant complications in practical calculations, in particular
at higher perturbative orders.
Due to these complications, in this letter we refrain from the explicit
evaluation of $\gamma_5^{\AC}$-odd traces. Instead, in Sec.\,\ref{sec:acG5uv}
we show how this can be avoided at the practical level.

\subsubsection*{FDF}

Finally we evaluate the triangle diagrams by utilizing the \FDF\
approach. Using the same four-dimensional Feynman rules as in
Sec.\,\ref{sec:oneLoopExample1}, the analytical expression reads
\begin{align}
 q_\mu\,T^{\mu\alpha\beta}_{\text{\A\V\V}}
 &\to\int\!\frac{d^{\dim}k}{(2\pi)^{\dim}}\frac{
 \text{Tr}\big[
  \slashed{q}
  \gamma_5^{}\,
  \big(
    \slashed{k}
    \!+\!i\,\mu\,\gamma_5^{}
    \!+\!\slashed{p}_{1}\big)\,
  \gamma^{\alpha}\,
  \big(
    \slashed{k}
    \!+\!i\,\mu\,\gamma_5^{}\big)\,
  \gamma^{\beta}\,
  \big(
    \slashed{k}
    \!+\!i\,\mu\,\gamma_5^{}
    \!-\!\slashed{p}_{2}\big)
    \big]_{[4]}
  }{
  (k\!+\!p_{1})^2_{[\dim]}\,
  k^2_{[\dim]}\,
  (k\!-\!p_{2})^2_{[\dim]}
  }
  \!+\!\bigg(\begin{aligned}
    p_1&\!\leftrightarrow\!p_2\\
    \alpha&\!\leftrightarrow\!\beta
   \end{aligned}\bigg).
  \label{eq:qTFDF}
\end{align}
A crucial difference compared to other dimensional schemes is the
appearance of rank two tensor integrals with strictly four-dimensional
loop momenta in the numerator,
\begin{subequations}
\begin{align}
  &\int\!\frac{d^{\dim}k}{(2\pi)^{\dim}}
    \frac{k_{[4]}^{\rho}\,k_{[4]}^{\sigma\phantom{\rho}}
  }{
  (k\!+\!p_{1})^2_{[\dim]}\,
  k^2_{[\dim]}\,
  (k\!-\!p_{2})^2_{[\dim]}
  }
  \equiv
    C_{00}\
      g^{\rho\sigma}_{[4]}
    +C_{12}\,\big(
      p_{1}^{\rho}\,p_{2}^{\sigma}
      \!+\!p_{2}^{\rho}\,p_{1}^{\sigma}
      \big)_{[4]}
    +\dots
  \,.
\end{align}
Using Eq.\,\eqref{eq:FDFshift} and neglecting odd powers of $\mu$,
the relevant coefficient is given by
\begin{align}
 C_{00}=\frac{1}{2}
 \bigg\{
 \int\!\frac{d^{\dim}k}{(2\pi)^{\dim}}
  \frac{1}{
  (k\!+\!p_{1})^2_{[\dim]}
  (k\!-\!p_{2})^2_{[\dim]}}
 +\int\!\frac{d^{\dim}k}{(2\pi)^{\dim}}
  \frac{k^2_{[4]}}{
  (k\!+\!p_{1})^2_{[\dim]}\,
   k^2_{[\dim]}\,
  (k\!-\!p_{2})^2_{[\dim]}}
  \bigg\}\,.
\end{align}
\end{subequations}
The first integrand is given by $\dim$-dimensional quantities only and the
integral can be evaluated without any complication. In contrast, the second
integral contains strictly four-dimensional components of the loop momentum.
Using Eq.\,\eqref{eq:FDFshift} to cancel against the denominator gives
rise to the integral in Eq.\,\eqref{eq:muInt} for $m\!=\!0$.
It turns out that this integral is the only one that contributes to the
AVV correlator in the \FDF\ approach. In other words, the anomaly is
entirely given by a $\mu^2$ integral that stems from the evaluation of
the tensor integrals,
\begin{subequations}
\begin{align}
 q_\mu\,T^{\mu\alpha\beta}_{\text{\A\V\V}}
 &\to 16i\, I_{3}^{\dim}(\mu^2)\,\Big\{
    \varepsilon^{\alpha\beta\mu\nu}\,p_{1,\mu}\,p_{2,\nu}
    \Big\}_{[4]}\\*
 &=-\frac{1}{2\pi^2}\,\Big\{
    \varepsilon^{\alpha\beta\mu\nu}\,p_{1,\mu}\,p_{2,\nu}
    \Big\}_{[4]}\Big[\,1\!+\!3\epsilon\!+\!\mathcal{O}(\epsilon^2)\Big]\,.
 \label{eq:axialAnomFDF}
\end{align}
\end{subequations}
In this way, the result in Eq.\,\eqref{eq:qTCDRBM} is recovered,
including higher terms in the $\epsilon$ expansion. Again, the
computational effort is significantly reduced compared to the
case of $\gamma_5^{\HVBM}$.

\subsubsection*{Comment on Bose symmetry}

Recently it has been shown \cite{Porto:2017asd,Viglioni:2016nqc}
that special care has to be taken when using an anticommuting $\gamma_5^{\AC}$
since gauge invariance and Bose symmetry may not be maintained simultaneously,
even if the dimension of the underlying space-time remains unchanged during
the regularization process. At the root of this symmetry breaking are
$\gamma_5^{\AC}$-odd traces which yield different contributions compared
to the case of $\gamma_5^{\HVBM}$.

In Ref.\,\cite{Viglioni:2016nqc}, the interplay between gauge invariance
and Bose symmetry is investigated in the framework of implicit regularization
(\IREG). Using $\gamma_5^{\HVBM}$ as defined in Eq.\,\eqref{eq:g5hvbm} together
with the right- and left-handed chiral operators
$V^{\mu}_{\text{\R}}
\!\equiv\!\frac{1}{2}\,\gamma^{\mu}_{\phantom{\text{\R}}}
(\,\mathbb{I}+\!\gamma_5^{\text{\HVBM}})$
and
$V^{\mu}_{\text{\L}}
\!\equiv\!\frac{1}{2}\,\gamma^{\mu}_{\phantom{\text{\R}}}
(\,\mathbb{I}-\!\gamma_5^{\text{\HVBM}})$
at the vertices, the following results for the different correlators are provided%
\footnote{In Ref.\,\cite{Viglioni:2016nqc}, the results are parametrized in
  terms of a parameter $a$ which is related to momentum-routing
  invariance and therefore to a so-called 'surface term' $v_0\!\sim\!(1\!+\!a)$.
  In dimensional regularization, $v_0$ is set to zero by definition, resulting
  in $a\!=\!-1$.}
\begin{subequations}
\label{eq:chiralBM}
 \begin{align}
 \underline{\text{\IREG}/\HVBM}:\qquad\qquad\qquad
 q_{\mu}\,T^{\mu\alpha\beta}_{\text{\R\R\R}}
 =-q_{\mu}\,T^{\mu\alpha\beta}_{\text{\L\L\L}}
 &=-\frac{1}{12\pi^2}\,\Big\{
    \varepsilon^{\alpha\beta\mu\nu}\,p_{1,\mu}\,p_{2,\nu}
    \Big\}_{[4]}
    \,,\\
  q_{\mu}\,T^{\mu\alpha\beta}_{\text{\R\R\L}}
 =q_{\mu}\,T^{\mu\alpha\beta}_{\text{\R\L\R}}
 =\frac{1}{2}q_{\mu}\,T^{\mu\alpha\beta}_{\text{\R\L\L}}
 &=-\frac{1}{24\pi^2}\,\Big\{
    \varepsilon^{\alpha\beta\mu\nu}\,p_{1,\mu}\,p_{2,\nu}
    \Big\}_{[4]}
    \,.
 \end{align}
\end{subequations}
In contrast, the same correlators read for the case of an anticommuting
$\gamma_5^{\AC}$
\begin{subequations}
\label{eq:chiralAC}
 \begin{align}
 \underline{\text{\IREG}/\AC}:\qquad\qquad\qquad
 q_{\mu}\,T^{\mu\alpha\beta}_{\text{\R\R\R}}
 =-q_{\mu}\,T^{\mu\alpha\beta}_{\text{\L\L\L}}
 &=-\frac{1}{12\pi^2}\,\Big\{
    \varepsilon^{\alpha\beta\mu\nu}\,p_{1,\mu}\,p_{2,\nu}
    \Big\}_{[4]}
    \label{eq:chiralACa}
    \,,\  \\
  q_{\mu}\,T^{\mu\alpha\beta}_{\text{\R\R\L}}
 =q_{\mu}\,T^{\mu\alpha\beta}_{\text{\R\L\R}}
 =q_{\mu}\,T^{\mu\alpha\beta}_{\text{\R\L\L}}
 &=0\,.
 \label{eq:chiralACb}
 \end{align}
\end{subequations}
The crucial difference between these two results is that only
Eqs.\,\eqref{eq:chiralBM} are likewise compatible with gauge invariance and
Bose symmetry since in this case Bose symmetry does not impose any additional
restrictions on the distribution of the anomaly on the pseudo-scalar and the
vector current~\cite{Viglioni:2016nqc}. It is therefore possible to entirely shift
the anomaly away from the vector current in order to preserve gauge invariance.

Using the \FDF\ approach, we computed the aforementioned chiral correlators
and find agreement with Eqs.\,\eqref{eq:chiralBM}, i.\,e.\
\begin{subequations}
\label{eq:chiralFDF}
 \begin{align}
 \underline{\text{\FDF}}:\qquad\qquad\qquad
 q_{\mu}\,T^{\mu\alpha\beta}_{\text{\R\R\R}}
 =-q_{\mu}\,T^{\mu\alpha\beta}_{\text{\L\L\L}}
 &=-\frac{1}{12\pi^2}\,\Big\{
    \varepsilon^{\alpha\beta\mu\nu}\,p_{1,\mu}\,p_{2,\nu}
    \Big\}_{[4]}%
    +\mathcal{O}(\epsilon)
    \,,\\
  q_{\mu}\,T^{\mu\alpha\beta}_{\text{\R\R\L}}
 =q_{\mu}\,T^{\mu\alpha\beta}_{\text{\R\L\R}}
 =\frac{1}{2}q_{\mu}\,T^{\mu\alpha\beta}_{\text{\R\L\L}}
 &=-\frac{1}{24\pi^2}\,\Big\{
    \varepsilon^{\alpha\beta\mu\nu}\,p_{1,\mu}\,p_{2,\nu}
    \Big\}_{[4]}%
    +\mathcal{O}(\epsilon)
    \,.
 \end{align}
\end{subequations}
In \FDF, the results are entirely generated by extra-integrals like in
Eq.\,\eqref{eq:muInt}. Although using a strictly four-dimensional algebra
in combination with an anticommuting $\gamma_5$, \FDF\ is therefore
compatible with Bose symmetry and gauge invariance at the same time.
This finding is confirmed by the validity of the vector Ward identities
for which we find in \FDF
\begin{align}
\underline{\text{\FDF}}:\qquad
 p_{1,\alpha}\,T^{\mu\alpha\beta}_{\text{\A\V\V}}
=p_{2,\beta}\,T^{\mu\alpha\beta}_{\text{\A\V\V}}=0\,.
\label{eq:VWIFDF}
\end{align}
It should be mentioned explicitly that these findings are a result of the
algebraic rules \textit{within} \FDF. If we were to evaluate the algebra
in the unregularized theory and apply the rules of \FDF\ only afterwards,
we would obtain vanishing results for the 'mixed' correlators
$\text{\R\R\L, \R\L\R, \R\L\L}$ like in Eq.\,\eqref{eq:chiralAC}
(although Eqs\,\eqref{eq:axialAnomFDF} and \eqref{eq:VWIFDF} would still
hold). Since the analytical expressions are in general divergent, however,
it is clear that the application of a proper regularization has to be the
initial step that is necessary to avoid ambiguous results.
\\

\section{Pseudo-scalar form factors in FDH}
\label{sec:psFF}

In the following, we extend the previous findings to the two-loop level by
computing the pseudo-scalar form factors of quarks and gluons in the \FDH\
scheme.
The results of the form factors that are currently available have been obtained
by using \CDR\ and $\gamma_5^{\HVBM}$ as defined in Eq.\,\eqref{eq:g5hvbm},
see e.\,g.\ \cite{Ahmed:2015qpa} and references therein. In the following we
consider the form factors up to two loops for
\begin{itemize}
 \item different dimensional schemes,
  i.\,e.\ \CDR/\HV\ and \FDH, and
 \item different $\gamma_5$ schemes,
  i.\,e.\ $\gamma_5^{\HVBM}$ and $\gamma_5^{\AC}$.
\end{itemize}
In principle, also the \FDF\ scheme is a viable candidate for treating
$\gamma_5$ in the framework of dimensional regularization. However, since
it is (currently) unclear how this approach can be consistently formulated
beyond the one-loop level, we do not consider \FDF\ here.

\subsection{Effective Lagrangian}
\label{sec:effL}

The coupling strength of a pseudo-scalar Higgs boson $A$ to quarks is directly
proportional to the respective quark mass. Denoting the pseudo-scalar current
by $j_{5,k}\!\equiv\!i\,\overline{\psi}_{k}\gamma_5\psi_{k}$, the corresponding
Lagrangian can be written as
\begin{align}
 \mathcal{L}_{\text{full}}=
 \Big[\,
  \sum_{q}\,y_{q}\,m_q\,j_{5,q}
  +y_{t}\,m_t\,j_{5,t}\,
  \Big]\,\frac{A}{v}\,,
 \label{eq:Lfull}
\end{align}
where $v$ and $y_{i}$ denote the Higgs vacuum expectation value and dimensionless
Yukawa couplings which depend on the underlying theory, respectively, the sum
runs over all light quark flavors $q\in\{d,u,s,c,b\}$, and $t$ corresponds to
the top quark.

One way to obtain an effective Lagrangian corresponding to Eq.\,\eqref{eq:Lfull}
is to consider the (all-order) anomalous relation \cite{Adler:1969er} between
the pseudo-scalar current $j_{5,k}$ and the axial-vector current
$j_{5,k}^{\mu}\equiv\overline{\psi}_{k}\gamma^{\mu}\gamma_5\psi_{k}$
in the full theory,
\begin{align}
 \partial_{\mu}\,\Big[
  \sum_q j_{5,q}^{\mu}
  +j_{5,t}^{\mu}\,
  \Big]
 =2 \Big[
  \sum_q m_{q}\,j_{5,q}
  +m_{t}\,j_{5,t}
  \Big]
  +\frac{N_F\!+\!1}{2}\Big(\frac{\alphas}{4\pi}\Big)
 \varepsilon^{\mu\nu\rho\sigma}G_{\mu\nu}^{a}G_{\rho\sigma}^{a} \, ,
 \label{eq:AB}
\end{align}
where  $G_{\mu\nu}^a$ is the gluonic field strength tensor and
$\alphas\!=\!g_s^2/(4\pi)$ denotes the strong coupling.
In the limit of a large top mass, $m_t^2\!\gg\!p^2$, the derivative
$\partial_{\mu}\,j_{5,t}^{\mu}$ and the masses of the light quarks
can be neglected.
The (unregularized) effective Lagrangian can then be written as~%
\cite{
Chetyrkin:1998mw}
\begin{align}
 \mathcal{L}_{\text{eff}}=
 \bigg[
  -\frac{\lambda_{\G}}{8}\,\Big\{
    \varepsilon^{\mu\nu\rho\sigma}\,G_{\mu\nu}^{a}G_{\rho\sigma}^{a}\Big\}_{[4]}
  -\frac{\lambda_{\J}^{\phantom{I}}}{2}\,\Big\{
  \partial_{\mu}\,\big(\,\sum_q
  \overline{\psi}_{q}\,\gamma^{\mu}\gamma_{5}\,\psi_{q}\,\big)\Big\}_{[4]}
  \bigg]\,A\,,
  \label{eq:Leff}
\end{align}
where the $\psi_q$ are now quark fields in the effective theory. One
important feature of the effective Lagrangian is that it does not carry
any mass dependence anymore. Although the interaction between a
(pseudo-scalar) Higgs and quarks vanishes in the full theory if the quark
masses are set to zero, in the effective theory we consider the case of
$N_F$ massless quarks which are described by the field $\psi$.
The implications of this choice will be discussed below.

In a next step, we study the effective Lagrangian~\eqref{eq:Leff} in the
framework of the aforementioned dimensional schemes. For this, it is useful
to envision some universal characteristics of dimensionally regularized
quantities. In any dimensional scheme, derivatives and loop momenta are
treated as (quasi) $\dim$-dimensional objects. In contrast, for the
dimensionality of metric tensors, $\gamma$ matrices, and vector fields
there is some freedom which is fixed by the choice of a specific
regularization scheme. In \CDR, for example, all Lorentz indices
(except for the ones of the $\varepsilon$ pseudotensor) are treated in
$\dim$ dimensions. The \CDR-regularized version of the first
curly bracket in Eq.\,\eqref{eq:Leff} therefore reads
\begin{subequations}
\label{eq:OG}
\vspace{-1pt}
\begin{align}
 O_{\G,\,\text{\CDR}} &\equiv\Big\{\varepsilon^{\mu\nu\rho\sigma}\Big\}_{[4]}\,
 \Big\{G_{\mu\nu}^{a}G_{\rho\sigma}^{a}\Big\}_{[\dim]}\, .
 \label{eq:OGCDR}
\end{align}
The corresponding Feynman rules are given in Appendix~\ref{sec:FeynmanRules}.

One key feature of the Feynman rules stemming from operator~\eqref{eq:OGCDR}
is that all of them contain (quasi) $\dim$-dimensional momenta with
uncontracted Lorentz indices. Due to permutations in $\mu,\nu,\rho,\sigma$,
the metric tensors in Eqs.\,\eqref{frO3a} and \eqref{frO4a} also have to be
considered in $\dim$ dimensions. The dimensionality of the indices
in Eq.\,\eqref{eq:OGCDR} is therefore valid in \textit{all} realizations
of dimensional regularization, i.\,e.\
\vspace{-1pt}
\begin{align}
 O_{\G,\,\text{\CDR}}=
 O_{\G,\,\text{\HV}}=
 O_{\G,\,\text{\FDH}}=
 O_{\G,\,\text{\DRED}}
 \equiv O_{\G}\,.
 \label{eq:OGDREG}
\end{align}
\end{subequations}
This in particular means that in \FDH\ and \DRED\ no evanescent operators
related to $\epsilon$-scalar--Higgs interactions arise at the tree level.

The regularization of the second curly bracket in Eq.\,\eqref{eq:Leff} is more
involved due to the treatment of $\gamma_5$. According to the discussion in
Sec.~\ref{sec:g5} we obtain the regularized operators%
\footnote{Eq.\,\eqref{eq:OJhv} is obtained by starting from the unregularized
  Lagrangian~\eqref{eq:Leff} and applying the shift of Eq.\,\eqref{eq:g5symm}
  together with Def.\,\eqref{eq:g5hvbm}. The structure of operator
  $O_{\J,\text{\CDR}}^{\HVBM}$ has first been discussed in
  Ref.\,\cite{Akyeampong:1973xi}. }
\begin{subequations}
\label{eq:OJreg}
\vspace{-1pt}
\begin{align}
 &\underline{\HVBM}:\qquad
 &&O_{\J,\text{\CDR}}^{\HVBM}
 =\frac{i}{3!}
  \Big\{\varepsilon^{\mu\nu\rho\sigma}\Big\}_{[4]}\,\,
  \Big\{\partial_{\mu}\big(\,
  \overline{\psi}\,\gamma_{\nu}\gamma_{\rho}\gamma_{\sigma}\,\psi\,
  \big)\Big\}_{[\dim]}\quad\text{and} &
  \label{eq:OJhv}
 \\*[.25cm]
 &\underline{\AC}:\qquad
 &&O_{\J,\text{\CDR}}^{\AC\,}
 = \Big\{
  \partial^{\mu}\,\big(\,
  \overline{\psi}\,\gamma_{\mu}\gamma_{5}^{\AC}\,\psi\,\big)\Big\}_{[\dim]}\,.&
 \label{eq:OJac}
\end{align}
\end{subequations}
In analogy to the discussion of operator $O_{\G}$ it follows that
Eqs.\,\eqref{eq:OJreg} are valid in \textit{all} implementations of
dimensional regularization,
\vspace{-1pt}
\begin{align}
 O_{\J,\,\text{\CDR}}^{\GS}
 =O_{\J,\,\text{\HV}}^{\GS}
 =O_{\J,\,\text{\FDH}}^{\GS}
 =O_{\J,\,\text{\DRED}}^{\GS}
 \equiv O_{\J}^{\GS}\,.
 \label{eq:OJDREG}
\end{align}
As for operator $O_{\G}$, the corresponding Feynman rules are given in
Appendix~\ref{sec:FeynmanRules}.\\

\subsection{Common definition of the form factors}
\label{sec:psFFcommon}

The regularized operators in Eqs.\,\eqref{eq:OG} and \eqref{eq:OJreg}
give rise to different pseudo-scalar form factors of quarks and gluons.
So far, in the literature these quantities have been considered in the
framework of \CDR, using $\gamma_5^{\HVBM}$ as defined in Eq.\,\eqref{eq:g5hvbm}.
The quark form factor related to contributions from operator~\eqref{eq:OJhv},
for example, is usually defined via squares of the absolute value of the
corresponding matrix elements,
\begin{align}
  f_{q,\J}^{\,\HVBM}
  &\equiv\sum_{n=0}^{\infty}
  \frac{
  \langle\,M^{\HVBM,(0)}_{q,\J}\,|\,
  M^{\HVBM,(n)}_{q,\J}\,\rangle}{
  \langle\,M^{\HVBM,(0)}_{q,\J}\,|\,
   M^{\HVBM,(0)}_{q,\J}\,\rangle}
   \equiv1+f_{q,\J}^{\,\HVBM,(1)}+f_{q,\J}^{\,\HVBM,(2)}
     +\mathcal{O}(\alpha_s^3)\, ,
   \label{eq:bareFFaOLD}
\end{align}
where $n$ denotes the loop order in the perturbative expansion. By definition,
each term in the sum contains products of two $\varepsilon$ pseudotensors.
Although the $\varepsilon^{\mu\nu\rho\sigma}$ are strictly four-dimensional
objects, in the literature their products are usually treated in $\dim$
dimensions~\cite{Zijlstra:1992kj},%
\footnote{In order to distinguish this $\dim$-dimensional treatment of the
  $\varepsilon$ pseudotensor from a strictly four-dimensional one we use
  the symbol $E$.}
\begin{align}
 \Big\{
  E^{\mu_1\mu_2\mu_3\mu_4}
  E^{\nu_1\nu_2\nu_3\nu_4}
 \Big\}_{[d]}
 \equiv
 \Big\{
 -g^{\mu_1\nu_1}g^{\mu_2\nu_2}g^{\mu_3\nu_3}g^{\mu_4\nu_4}
  \pm\text{perm.}
 \Big\}_{[d]}\,,
 \label{eq:epsSqD}
\end{align}
where 'perm.' denotes terms originating from further permutations in the
Lorentz indices. Even though the application of Eq.\,\eqref{eq:epsSqD} in
general leads to ambiguous results~\cite{Siegel:1980qs}, we consider the
implications of this choice by using it to evaluate the numerators in
Eq.\,\eqref{eq:bareFFaOLD}.

If $p_1,\,p_2$ denote the momenta of the external quarks with
$q\equiv p_1\!+p_2$ and $p_1^2=p_2^2\equiv p^2$, we obtain for the
first numerator in the perturbative expansion%
\begin{align}
\langle\,M^{\HVBM,(0)}_{q,\J}\,|\,
  M^{\HVBM,(0)}_{q,\J}\,\rangle
 &= \Big(\frac{i}{3!}\Big)^2
 q_{\mu_1}\,q_{\nu_1}
 \Big\{
  E^{\mu_1\mu_2\mu_3\mu_4}
  E^{\nu_1\nu_2\nu_3\nu_4}
 \Big\}_{[d]}
 \Big\{
  \text{Tr}\Big[
  \gamma_{\mu_4}\gamma_{\mu_3}\gamma_{\mu_2}
  \,\slashed{p}_1\,
  \gamma_{\nu_2}\gamma_{\nu_3}\gamma_{\nu_4}
  \,\slashed{p}_2
  \Big]
 \Big\}_{[d]}
 \notag\\*
 &= \!-\frac{1}{3}\,q^2\,\Big[q^2\,\big(\dim-4\big)+p^2\,(14-2\dim)\Big]
  \big(\dim-3\big)\big(\dim-2\big) \, .
 \label{eq:treeSqHV}
\end{align}
It follows that in the massless on-shell case ($p^2\!=\!0$), the use of
Eq.\,\eqref{eq:epsSqD} serves as an intermediate regularization of the
fractions in Eq.\,\eqref{eq:bareFFaOLD}. This regularization has to be
introduced since the r.\,h.\,s.\ of Eq.\,\eqref{eq:treeSqHV} vanishes for
$\dim=4$.
Since the regulator drops out in the definition of the form factors, however,
the effects of vanishing quark masses are eliminated. In this way, the Lorentz
structure related to the pseudo-scalar vertex is effectively disentangled from
the kinematics of the process and only the (anti)commutation property of the
$\varepsilon$ pseudotensor is kept.

In contrast, a separation between the Lorentz structure and the mass
dependence of the effective Lagrangian is not possible when using an
anticommuting $\gamma_5^{\AC}$. In this case, the square of the absolute
value vanishes in the massless on-shell case, even for arbitrary $\dim$,
\begin{align}
\langle\,M^{\AC,(0)}_{q,\J}\,|\,M^{\AC,(0)}_{q,\J}\,\rangle
&\ \sim\
  \text{Tr}\Big[
    \gamma_5^{\AC}\,
    \slashed{q}\,
    \slashed{p}_{1}\,
    \slashed{q}\,
    \gamma_5^{\AC}\,
    \slashed{p}_{2}
  \Big]
 = -4\,q^2 p^2 \, .
 \label{eq:treeSqAC}
\end{align}
An on-shell definition of the form factor for the case of massless quarks
similar to the one in Eq.\,\eqref{eq:bareFFaOLD} is therefore \text{not}
possible for $\gamma_5^{\AC}$. The reason is that using Eq.\,\eqref{eq:g5ac}
as the defining property of $\gamma_5^{\AC}$, the $\varepsilon$ pseudotensor
is \textit{implicitly} treated in strictly four dimensions. Like in
Eq.\,\eqref{eq:treeSqHV} for $\dim\!=\!4$, the square of the tree-level
amplitudes then vanishes for $p^2\!=\!0$. In the next section we provide
alternative definitions of the pseudo-scalar form factors, avoiding the
use of Eq.\,\eqref{eq:epsSqD} and including the case of an anticommuting
$\gamma_5^{\AC}$.\\

\subsection{Alternative definition and bare results for $\gamma_5^{\text{BM}}$}
\label{sec:epstensor}

Like in Secs.\,\ref{sec:oneLoopExample1} and \ref{sec:oneLoopExample2},
in the following we consider the $\varepsilon$ pseudotensor outside
dimensional regularization and treat it in strictly four dimensions.
In this way it is only the remainder that is dimensionally regularized.
Following Ref.~\cite{Chetyrkin:1998mw}, we write, for example, the (all-order)
contribution of operator $O_{\J}^{\HVBM}$ to the quark form factor as%
\footnote{All other form factors involving $\varepsilon$ pseudotensors
  are treated in the same way. The corresponding definitions are given in
Appendix~\ref{sec:FeynmanRules}.}
 \begin{align}
  M_{q,\J}^{\HVBM}
  &=\Big\{\varepsilon_{\mu\nu\rho\sigma}\Big\}_{[4]}\,
    \bar{u}(p_1)\,\Big\{
    \big(R_{q,\J}^{\HVBM}\big)^{\mu\nu\rho\sigma}\Big\}_{[d]}\,
    v(p_2)\,,
    \label{eq:amp1}
 \end{align}
where $u$ and $v$ denote spinors of the external quarks. By construction,
the remainder in the second curly bracket is totally antisymmetric in
$\mu,\nu,\rho,\sigma$. Regarding its Lorentz decomposition there is only
one structure that is linear in the external momentum~$q$. Making the
(anti)symmetrization explicit, we write
\begin{align}
  \big(R_{q,\J}^{\HVBM}\big)^{\mu\nu\rho\sigma}
  &\ =\ 
    \big(
      q^{\mu}\gamma^{\nu}\gamma^{\rho}\gamma^{\sigma}
      \!-\!q^{\nu}\gamma^{\mu}\gamma^{\rho}\gamma^{\sigma}
      \pm \text{perm.}\big)\,
      R_{q,\J}^{\HVBM}
  \ \ \equiv\ 
  \big(P_{q}^{\HVBM}\big)^{\mu\nu\rho\sigma}\,
  R_{q,\J}^{\HVBM}\,.
  \phantom{\bigg|}
  \label{eq:amp2}
\end{align}
For the extraction of the remainder without indices
we define the normalization factor
\begin{align}
  \text{Tr}\Big[
    q_{\mu}\gamma_{\nu}\gamma_{\rho}\gamma_{\sigma}\,
    \big(P_{q}^{\HVBM}\big)^{\mu\nu\rho\sigma}
    \Big]_{[d]}
    &=\frac{q^2}{6}(d\!-\!3)\,(d\!-\!2)\,(d\!-\!1)
    \equiv N_{q}^{\HVBM}\,.
    \phantom{\bigg|}
\end{align}
The coefficient of the remainder is then obtained by
\begin{align}
  R_{q,\J}^{\HVBM}
  &=\big(N_{q}^{\HVBM}\big)^{-1}\,\text{Tr}\Big[
    q_{\mu}\gamma_{\nu}\gamma_{\rho}\gamma_{\sigma}\
    \big(R_{q,\J}^{\HVBM}\big)^{\mu\nu\rho\sigma}
    \Big]\, .
    \phantom{\bigg|}
 \label{eq:obtainR}
\end{align}
In practical calculations we directly implement the quantity
$\big(R_{q,\J}^{\HVBM}\big)^{\mu\nu\rho\sigma}$.
In other words, we use the Feynman rule in Eq.\,\eqref{frO1a} and suppress
the $\varepsilon$ pseudotensor. Using the projection in Eq.\,\eqref{eq:obtainR},
this modified Feynman rule is then used to compute one- and two-loop
contributions of operator $O_{\J}^{\HVBM}$ to the form factor.
In general, these results are UV and IR divergent. After UV renormalization
and IR subtraction, however, the limit $d\!\to\!4$ can be taken. A contraction
with the four-dimensional indices of $\varepsilon^{\mu\nu\rho\sigma}$ is
then possible.

Using this approach and $\gamma_{5}^{\HVBM}$ as given in Eq.\,\eqref{eq:g5hvbm},
we define the (regularization- scheme dependent) pseudo-scalar form factors of
quarks
\begin{subequations}
\label{eq:bareFF}
\begin{align}
  \bar{F}_{q,\J}^{\HVBM}
  &\equiv\sum_{n=0}^{\infty}\,
  \frac{\bar{R}^{\HVBM,(n)}_{q,\J}}{R^{\HVBM,(0)}_{q,\J}}
  \equiv1+\bar{F}_{q,\J}^{\HVBM,(1)}+\bar{F}_{q,\J}^{\HVBM,(2)}
    +\mathcal{O}(\alpha_s^3)\, ,
  \label{eq:bareFFa}
  \\*
  \bar{F}_{q,\G}
  &\equiv\sum_{n=1}^{\infty}\,
  \frac{\bar{R}^{(n)}_{q,\G}}{R^{(1)}_{q,\G}}
  \equiv1+\bar{F}_{q,\G}^{(1)}
    +\mathcal{O}(\alpha_s^2) \, ,
\end{align}
and gluons
\begin{align}
  \bar{F}_{g,\G}
  &\equiv\sum_{n=0}^{\infty}
  \frac{\bar{R}^{(n)}_{g,\G}}{R^{(0)}_{g,\G}}
  \equiv1+\bar{F}_{g,\G}^{(1)}+\bar{F}_{g,\G}^{(2)}
    +\mathcal{O}(\alpha_s^3)\, ,
  \\*
  \bar{F}_{g,\J}^{\HVBM}
  &\equiv\sum_{n=1}^{\infty}
  \frac{\bar{R}^{\HVBM,(n)}_{g,\J}}{R^{\HVBM,(1)}_{g,\J}}
  \equiv1+\bar{F}_{g,\J}^{\HVBM,(1)}
    +\mathcal{O}(\alpha_s^2)\, .
\end{align}
\end{subequations}
The notation $\bar{R}_{a,\A}^{(n)}$ for the remainders is chosen such that
the index $n$ denotes the loop order in the perturbative expansion,
$a\!\in\!\{q,g\}$ indicates a contribution to the quark or the gluon form
factor, and $\A\!\in\!\{\J,\G\}$ specifies whether the respective contribution
originates from operator $O_{\J}$ or $O_{\G}$. The explicit definition of the
remainders is given in Appendix~\eqref{sec:FeynmanRules}.
To distinguish the underlying regularization we use a bar for quantities
in the \FDH\ scheme and no bar for quantities in \CDR/\HV. Note that
contributions related to operator $O_{\J}$ depend on the applied
$\gamma_5$ scheme which is indicated by the superscript $\HVBM$.%

\begin{figure}[t]
\begin{center}
\scalebox{.75}{
\begin{picture}(135,90)(0,10)
\DashLine(0,45)(30,45){2}
\Line[arrow](30,45)(120,90)
\Line[arrow](120,0)(30,45)
\Vertex(30,45){2}
\Text(20,60)[c]{\scalebox{1.33}{$O_{\J}^{\GS}$}}
\end{picture}
\quad
\begin{picture}(135,90)(0,10)
\DashLine(0,45)(30,45){2}
\Gluon(30,45)(100,80){4}{9}
\Line[arrow](100,80)(120,90)
\Line[arrow](120,0)(100,10)
\Gluon(30,45)(100,10){4}{9}
\Line[arrow](100,10)(100,80)
\Vertex(30,45){2}
\Vertex(100,80){2}
\Vertex(100,10){2}
\Text(20,60)[c]{\scalebox{1.33}{$O_{\G}$}}
\end{picture}
\quad
\begin{picture}(135,90)(0,10)
\DashLine(0,45)(30,45){2}
\Gluon(30,45)(120,90){4}{11}
\Gluon(30,45)(120,0){4}{11}
\Vertex(30,45){2}
\Text(20,60)[c]{\scalebox{1.33}{$O_{\G}$}}
\end{picture}
\quad
\begin{picture}(135,90)(0,10)
\DashLine(0,45)(30,45){2}
\Line[arrow](30,45)(100,10)
\Line[arrow](100,10)(100,80)
\Line[arrow](100,80)(30,45)
\Gluon(100,80)(120,90){4}{2}
\Gluon(100,10)(120,0){4}{2}
\Vertex(30,45){2}
\Vertex(100,80){2}
\Vertex(100,10){2}
\Text(20,60)[c]{\scalebox{1.33}{$O_{\J}^{\GS}$}}
\end{picture}
}
\end{center}
\caption{\label{fig:fig3}
Lowest-order contributions to the pseudo-scalar form factors
$\bar{F}_{q,\J}^{\GS}$, $\bar{F}_{q,\G}$, $\bar{F}_{g,\G}$,
and $\bar{F}_{g,\J}^{\GS}$ (from left to right).
Note, that the 'mixed' amplitudes $\bar{M}_{q,\G}$ and
$\bar{M}_{g,\J}$ are loop induced and are therefore at least of
$\mathcal{O}(\alphas)$.}
\end{figure}
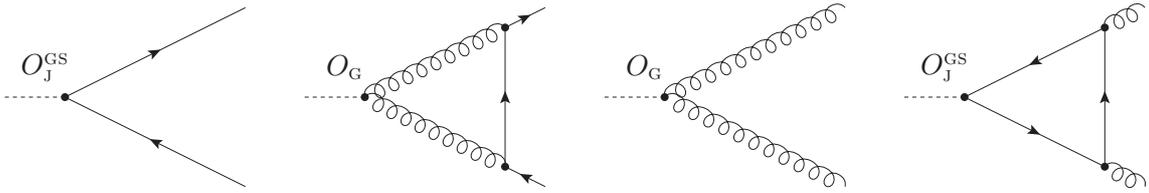

The lowest-order contributions to the form factors are shown in
Fig.\,\ref{fig:fig3}. As discussed in Sec.~\ref{sec:effL}, they do not
depend on the applied version of dimensional regularization, i.\,e.\
$\bar{R}^{\HVBM,(0)}_{q,\J}\!=\!R^{\HVBM,(0)}_{q,\J}$,
$\bar{R}^{(1)}_{q,\G}\!=\!R^{(1)}_{q,\G}$
and similar for amplitudes with external gluons. At higher perturbative
orders, however, \FDH\ results differ from the ones in \CDR/\HV\ due to
the different treatment of the Lorentz algebra.
For the practical calculations in the \FDH\ scheme we follow the guideline
given in Sec.\,4 of Ref.~\cite{Gnendiger:2016cpg}. More precisely, at the
one-loop level we perform the split of Eq.\,\eqref{eq:vsDecomp} and
distinguish the evanescent coupling $\alphae$ which is related to
$\epsilon$-scalar--fermion interactions from the gauge coupling $\alphas$.%
\footnote{For the definition of $\alphae$ we refer to
  Ref.~\cite{Gnendiger:2014nxa}. The only one-loop diagram $\sim\!\alphae$
  that is relevant for the present computation is the right one in
  Fig.\,\ref{fig:fig5}.}
The two-loop calculations are performed by using a (quasi) $d_s$-dimensional
Lorentz algebra as given in Eqs.\,\eqref{eq:FDHalg}. Throughout the
calculation, $d_s$ is identified with $4$.

The one- and two-loop results of the (bare) form factors in \FDH\ are given
in Appendix~\ref{sec:bareFF}. They have been obtained in the following way:
The generation of the diagrams and analytical expressions has been done
with the Mathematica package FeynArts~\cite{Hahn:2000kx}. In order to cope
with the Lorentz structure in the \FDH\ scheme we used a modified
version of TRACER~\cite{Jamin:1991dp}. The subsequent integral reduction and
evaluation has been done with an in-house-algorithm that is based on
integration-by-parts identities and the Laporta algorithm~\cite{Laporta:2001dd}.\\

\subsection{Form factors with $\gamma_5^{\text{AC}}$}
\label{sec:FFg5ca}

As shown in Sec.\,\ref{sec:oneLoopExample1}, the evaluation of the Lorentz
algebra using an anticommuting $\gamma_5^{\AC}$ may lead to much simpler
analytical expressions compared to the case of $\gamma_5^{\HVBM}$. Since,
for example, one-loop contributions of operator $O_{\J}^{\AC}$ to the quark
form factor do not contain traces with $\gamma_5^{\AC}$, the corresponding
amplitude can be written as
\begin{align}
 \bar{M}_{q,\J}^{\AC,(1)}
 =\bar{u}(p_1)\,
 \Big\{\slashed{q}\gamma_5^{\AC}\,\bar{R}_{q,\J}^{\AC,(1)}\Big\}_{[\dim]}\,
 v(p_2)\,.
\end{align}
Suppressing the spinors and using $(\gamma_5^{\AC})^2\!=\!\mathbb{I}_{}$,
the remainder can be extracted via
\begin{align}
 \bar{R}_{q,\J}^{\AC,(1)}=\frac{1}{4q^2}\text{Tr}
  \Big[\gamma_5^{\AC}\slashed{q}\,\bar{M}_{q,\J}^{\AC,(1)}\Big]_{[\dim]}\,.
\end{align}
This remainder can be used to define a form factor in a
similar way as in Eq.\,\eqref{eq:bareFFa}. As it turns out, however, all
perturbative coefficients of the remainder vanish in the massless on-shell
case. This can be seen from the explicit analytical expression
\begin{align}
\bar{M}_{q,\J}^{\AC,(1)}\sim
 \int\!\frac{d^\dim k}{(2\pi)^\dim}\frac{
  \gamma^{\alpha}_{[\dim_s]}\,\big[
  (\slashed{k}\!+\!\slashed{p}_1)\,
  \slashed{q}\gamma_5^{\AC}\,
  (\slashed{k}\!-\!\slashed{p}_2)\,
  \big]_{[\dim]}\,
  \gamma^{\beta}_{[\dim_s]}\,
  \big(g_{\alpha\beta}\big)_{[\dim_s]}
  }{
  (k\!+\!p_1)^2_{[\dim]}\,
  (k\!-\!p_2)^2_{[\dim]}\,
  k^2_{[\dim]}
  }
 \,.
  \label{eq:ampAC}
\end{align}
Anticommuting $\gamma_5^{\AC}$ to the left, the evaluation of the
algebra only yields integrals that are scaleless for $p_1^2\!=\!p_2^2\!=\!0$.
Like in Eq.\,\eqref{eq:treeSqAC}, a separation between the Lorentz structure
and the mass dependence of the effective Lagrangian is then not possible.
An anticommuting $\gamma_5^{\AC}$ can therefore not be used to obtain the
quark form factor related to Lagrangian~\eqref{eq:Lfull} in a massless
framework.%
\footnote{The fact that the amplitude vanishes for $\gamma_5^{\AC}$
  is not a characteristic of the $\AC$ scheme itself but of the observable
  under consideration. Even using $\gamma_5^{\HVBM}$, the square of the
  absolute values in Eq.\,\eqref{eq:treeSqHV} vanishes in the massless
  on-shell case if the $\varepsilon$ pseudotensors are treated in strictly
  four dimensions.}
However, in Sec.~\ref{sec:acG5uv} we consider Eq.\,\eqref{eq:ampAC} in the
massless off-shell case to determine so far unknown UV renormalization
constants. In this case, the amplitude has a non-vanishing value.\\

\section{UV renormalization}
\label{sec:uv}

\begin{figure}[t]
\begin{center}
\scalebox{.75}{
\begin{picture}(135,90)(0,10)
\end{picture}
\quad
\begin{picture}(135,90)(0,10)
 \DashLine(0,45)(30,45){2}
 \Line[arrow](30,45)(100,80)
 \Line[arrow](100,80)(120,90)
 \Line[arrow](120,0)(100,10)
 \Line[arrow](100,10)(30,45)
 \Gluon(100,10)(100,80){4}{8}
 \Vertex(30,45){2}
 \Vertex(100,80){2}
 \Vertex(100,10){2}
 \Text(20,60)[c]{\scalebox{1.33}{$O_{\J}^{\GS}$}}
\end{picture}
\quad
\begin{picture}(135,90)(0,10)
 \DashLine(0,45)(30,45){2}
 \Line[arrow](30,45)(100,80)
 \Line[arrow](100,80)(120,90)
 \Line[arrow](120,0)(100,10)
 \Line[arrow](100,10)(30,45)
 \DashLine(100,10)(100,80){4}
 \Vertex(30,45){2}
 \Vertex(100,80){2}
 \Vertex(100,10){2}
 \Text(20,60)[c]{\scalebox{1.33}{$O_{\J}^{\GS}$}}
\end{picture}
\quad
\begin{picture}(135,90)(0,10)
\end{picture}
}
\end{center}
\caption{\label{fig:fig5}
One-loop diagrams contributing to the form factor $\bar{F}_{q,\J}^{\GS}$
including a gluon (left) and an associated $\epsilon$-scalar (right).
The right diagram is proportional to the evanescent coupling $\alphae$
and only contributes in the \FDH\ scheme.}
\end{figure}
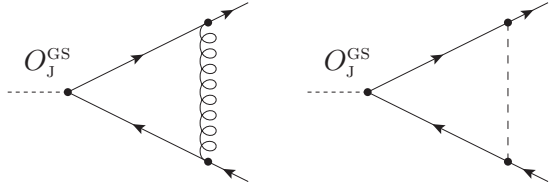

To obtain UV-renormalized Green functions it is useful to 
distinguish two classes of contributions,
\begin{itemize}
 \item renormalization of the couplings, fields, and the gauge parameter,
 \item renormalization of the effective operators
  $O_{\G}^{\phantom{\J}}$ and $O_{\J}^{\GS}$.
\end{itemize}
The renormalization of evanescent couplings in the \FDH\ scheme is well known%
~\cite{Harlander:2006rj, Harlander:2006xq}. In any $l$-loop calculation, the
coupling $\alphae$ describing the interaction of $\epsilon$-scalars and quarks
has to be distinguished from the gauge coupling $\alphas$ in $(l\!-\!1)$-loop 
contributions~\cite{Gnendiger:2016cpg}, see also Fig.\,\ref{fig:fig5}. The
multiplicative coupling renormalization is given by
\begin{align}
 \alpha_{i}^{0}\,=\,
 \Big(\frac{\mu_{r}}{\mu_{0}}\Big)^{2\epsilon}\,
   \bar{Z}_{\alpha_{i}}
   \,\alpha_{i}(\mu_{r})\, ,\qquad\quad
   \alpha_{i}^{0}\in\{\alpha_{s}^{0},\alpha_{e}^{0}\}\, ,
 \label{eq:couplingRen}
\end{align}
where $\mu_{r}$ and $\mu_{0}$ denote the renormalization scale and
the regularization scale, respectively. In the following we set
$\mu_{r}\equiv\mu_{0}$ and suppress the explicit scale dependence
of the renormalized couplings; as renormalization prescription we use
the $\MS$ scheme. The corresponding renormalization constants in
\FDH\ are given in Appendix~\ref{sec:UVrenMSbar}.

\subsection{Operator renormalization for $\gamma_5^{\text{BM}}$}
\label{sec:bmG5uv}

To describe the UV behavior of the operators $O_{\G}$ and $O_{\J}^{\GS}$,
multiplicative renormalization transformations similar to
Eq.\,\eqref{eq:couplingRen} are not sufficient since the operators mix under
renormalization.
As shown in Sec.\,\ref{sec:effL}, the operator basis remains unchanged when
using the \FDH\ scheme instead of \CDR\ due to the absence of evanescent
operators at the tree-level. The related renormalization constants, however,
are different in both schemes. In analogy to the \CDR\ result~\cite{Larin:1993tq},
we therefore write the operator mixing in \FDH\ as%
\footnote{Compared to the original reference we added the superscript
  $\GS$ indicating the dependence on the applied $\gamma_5$ scheme.
  The renormalization constants in \CDR\ are defined in the same way
  without a bar}
\begin{align}
 \left(\begin{aligned}
  &O_{\G}\\
  &O_{\J}
 \end{aligned}\right)_{\text{ren.}}
 \equiv
 \left(\begin{aligned}
  &\bar{Z}_{\G\G} 		&\bar{Z}_{\G\J}\\
  &\bar{Z}_{\J\G}^{\GS}		&\bar{Z}_{\J\J}^{\GS}
 \end{aligned}\right)
 \left(\begin{aligned}
  &O_{\G}\\
  &O_{\J}^{\GS}
 \end{aligned}\right)_{\text{bare}}.
 \label{eq:opmix}
\end{align}
The 'mixed' constants $\bar{Z}_{\G\J}$ and $\bar{Z}_{\J\G}^{\GS}$ are related to
UV divergences of the second and the rightmost diagram in Fig.\,\ref{fig:fig3},
respectively, and to perturbative corrections thereof.
As shown in Ref.\,\cite{Espriu:1982bw}, the latter constant vanishes to all
orders in perturbation theory, i.\,e.\ $\bar{Z}_{\J\G}^{\GS}=0$. The former,
on the other hand, is at least of $\mathcal{O}(\alphas)$.
Due to the absence of evanescent contributions to the second topology in
Fig.\,\ref{fig:fig3}, its one-loop coefficient is regularization-scheme
independent,
\begin{align}
  \bar{Z}_{\G\J}^{(1)}-Z_{\G\J}^{(1)}=0\,.
  \label{eq:dZGJ1}
\end{align}

As discussed in Sec.\,\ref{sec:oneLoopExample1}, the use of $\gamma_5^{\HVBM}$
in a dimensional framewrok spoils properties of the axial-vector current
and the Ward identities. In this case an additional \textit{finite}
renormalization $\bar{Z}_5^{\HVBM}$ has to be introduced to restore
the initial properties~\cite{Trueman:1979en}. We therefore define
\begin{align}
  \bar{Z}_{\J\J}^{\HVBM} \equiv \bar{Z}_{\MS}^{\HVBM}\,\bar{Z}_5^{\HVBM} \, ,
  \label{eq:ZJJdef}
\end{align}
where $\bar{Z}_{\MS}^{\HVBM}$ only contains pure poles in $\epsilon$ for
arbitrary $\Neps$. For the operator renormalization in the \FDH\ scheme
we then get
\begin{subequations}
\label{eq:opRen}
\begin{align}
 \big(O_{\G}\big)_{\text{ren.}}
 &= \bar{Z}_{\G\G}\ \big(O_{\G}\big)_{\text{bare}}
  +\bar{Z}_{\G\J}\ \big(O_{\J}^{\HVBM}\big)_{\text{bare}}\, ,
  \phantom{\bigg|}
  \label{eq:opRen1}
 \\*
 \big(O_{\J}\big)_{\text{ren.}}
 &= \bar{Z}_{\MS}^{\HVBM}\,\bar{Z}_5^{\HVBM}\,
    \big(O_{\J}^{\HVBM}\big)_{\text{bare}}\,.
 \label{eq:opRen2}
\end{align}
\end{subequations}
The values of $\bar{Z}_{\G\G}$, $\bar{Z}_{\G\J}$, and $\bar{Z}_{\MS}^{\HVBM}$
in the \FDH\ scheme can be obtained by making use of the fact that they are
the only so far unknown quantities entering the UV-renormalized and IR-subtracted
form factors. Using Eq.\,\eqref{eq:dZGJ1} and the structure of the IR divergences
given in Eqs.\,\eqref{eq:IRfac}, the particular structure of the operator
mixing allows one to determine the one- and two-loop renormalization coefficients
in a \textit{unique} way.

To illustrate the determination of the renormalization constants we consider
the renormalized form factor $\bar{\mathcal{F}}_{q,\J}^{\HVBM,(1)}$ given by
Eqs.\,\eqref{eq:FgJ1bare} and \eqref{eq:FFren1} as an example.
At the one-loop level, any UV renormalization constant has at most \textit{single}
$\epsilon$ poles in the framework of dimensional regularization. Depending on which
specific scheme is used, the coefficients of these poles differ by terms
$\sim\Neps$, depending on the treatment of metric tensors and $\gamma$ matrices.
The scheme-dependent part of a one-loop renormalization constant is therefore
finite for \mbox{$\Neps\!=\!2\epsilon$},
\begin{align}
 \Big(\delta\bar{Z}_{}^{(1)}-\delta Z_{}^{(1)}\Big)
 =\mathcal{O}(\Neps/\epsilon)
 =\mathcal{O}(\epsilon/\epsilon)
 =\mathcal{O}(\epsilon^0) \, .
  \phantom{\Big|}
\end{align}
In order to make the scheme-dependent terms explicit, however, we leave $\Neps$
as an arbitrary variable in the following results. Identifying the (renormalized)
couplings, $\alphae\!=\!\alphas$, a comparison of Eq.\,\eqref{eq:FFren1} with
prediction~\eqref{eq:IRprediction} for the IR divergences then yields 
\begin{subequations}
\begin{align}
 &\bar{\mathcal{F}}_{q,\J}^{\HVBM,(1)}-\mathcal{F}_{q,\J}^{\HVBM,(1)}
  \Big|_{\text{poles}}
 &&=\quad\ \ \Big(\frac{\alphas}{4\pi}\Big)
  \Big[-C_F\frac{\Neps}{2\,\epsilon}\Big]
 +\Big(\delta \bar{Z}_{\MS}^{\HVBM,(1)}-\delta Z_{\MS}^{\HVBM,(1)}\Big)
 &
 \\[5pt]
 \equiv\quad&
 \bar{\mathbf{Z}}^{(1)}_{q}
  -
  \mathbf{Z}^{(1)}_{q}
 &&=\quad\ \ \Big(\frac{\alphas}{4\pi}\Big)
  \Big[+C_F\frac{\Neps}{2\,\epsilon}\Big]\, .
  &
\end{align}
\end{subequations}
Since $\delta Z_{\MS}^{\HVBM,(1)}$ vanishes in \CDR ~\cite{Larin:1993tq},
$\bar{Z}_{\MS}^{\HVBM}$ receives a non-vanishing one-loop contribution
in the \FDH\ scheme which is finite for $\Neps=2\epsilon$,
\begin{align}
 \delta\bar{Z}_{\MS}^{\HVBM,(1)}
 =\Big(\frac{\alphas}{4\pi}\Big)\,
  C_F\frac{\Neps}{\epsilon}\, .
  \phantom{\bigg|}
\end{align}
All other renormalization coefficients can be obtained in
the same way. The explicit calculation yields
\begin{subequations}
\label{eq:ZopMixFDH}
\begin{align}
\bar{Z}_{\MS}^{\HVBM}&=1\!+\!
\Big(\frac{\alphas}{4\pi}\Big)\,C_F\frac{\Neps}{\epsilon}
+\Big(\frac{\alphas}{4\pi}\Big)^2 \Big\{ 
  C_A C_F \Big[
    \frac{22}{3\epsilon}
    +\Neps\Big(
      \!-\!\frac{1}{\epsilon^2}
      \!+\!\frac{11}{3\epsilon }
      \Big)
    +\Neps^2\Big(
      \frac{1}{2\epsilon^2}
      \!+\!\frac{1}{4\epsilon}
      \Big)
    \Big]
\notag\\ &\quad\ \
  +C_F^2\Big[
    \Neps\Big(
      \!-\!\frac{1}{\epsilon^2}
      \!-\!\frac{4}{\epsilon}
      \Big)
    -\frac{3\Neps^2}{4\epsilon}
    \Big]
  +C_F N_F \Big[
    \frac{5}{3\epsilon}
    \!+\!\Neps\Big(
      \frac{1}{2\epsilon ^2}
      \!-\!\frac{1}{4\epsilon}
      \Big)
      \Big]
    \Big\}
 +\mathcal{O}(\alphas^3)\,,
 \label{eq:ZmsHV}
\\[.35cm]
\bar{Z}_{\G\G} &=1\!+\!
\Big(\frac{\alphas}{4\pi}\Big)\Big\{
  C_A \Big[
    \!-\!\frac{11}{3\epsilon}
    \!+\!\frac{\Neps}{6\epsilon }
    \Big]
  \!+\!N_F\frac{2}{3\epsilon}
  \bigg\}
\!+\!\Big(\frac{\alphas}{4\pi}\Big)^2 \bigg\{
  C_A^2 \Big[
    \frac{121}{9\epsilon^2}
    \!-\!\frac{17}{3\epsilon }
    \!-\!\Neps\Big(
      \frac{11}{9\epsilon^2}
      \!-\!\frac{7}{6\epsilon}
      \Big)
    \!+\!\frac{\Neps^2}{36\epsilon^2} 
    \Big]
\notag\\ &\quad\ \
  +C_A N_F \Big[
    \!-\!\frac{44}{9\epsilon^2}
    \!+\!\frac{5}{3\epsilon}
    \!+\!\frac{2\Neps}{9\epsilon^2}
    \Big]
  +C_F N_F \Big[
    \frac{1}{\epsilon}
    \!-\!\frac{\Neps}{2\epsilon }
    \Big]
    +N_F^2\frac{4}{9\epsilon^2}
    \Big\}
\!+\mathcal{O}(\alphas^3)\,,
\\[.45cm]
\bar{Z}^{}_{\G\J}&=
\Big(\frac{\alphas}{4\pi}\Big)\,
  C_F\frac{12}{\epsilon }
+\Big(\frac{\alphas}{4\pi}\Big)^2 \Big\{ 
  C_A C_F \Big[
    \!-\!\frac{44}{\epsilon^2}
    \!+\!\frac{142}{3\epsilon}
    \!+\!\Neps\Big(
      \frac{2}{\epsilon^2}
      \!+\!\frac{2}{3\epsilon}
      \Big)
    \Big]
\notag\\ &\quad\ \
  +C_F^2\Big[
    \!-\!\frac{42}{\epsilon}
    \!+\!\Neps\Big(
      \frac{6}{\epsilon^2}
      \!-\!\frac{6}{\epsilon}
      \Big)
    \Big]
  +C_F N_F\Big[
    \frac{8}{\epsilon^2}
    \!-\!\frac{4}{3\epsilon}
    \Big]
  \Big\}
\!+\mathcal{O}(\alphas^3)\,.
\end{align}
\end{subequations}
For $\Neps\!=\!0$, Eqs.\,\eqref{eq:ZopMixFDH} agree with the well-known
\CDR\ results given e.\,g.\ in Ref.\,\cite{Larin:1993tq}. Like in \CDR,
$\bar{Z}_{\G\G}$ coincides with the renormalization of the gauge coupling,
see also Eq.\,\eqref{eq:Zalphas}.
The results in Eqs.\,\eqref{eq:ZopMixFDH} have been cross-checked with an
explicit calculation of the form factors in the off-shell case, including
a renormalization of the external parton fields and the gauge parameter.

The \CDR\ value of the \textit{finite} renormalization constant in
Eq.\,\eqref{eq:opRen2} is known up to the two-loop level~\cite{Larin:1993tq},
\begin{subequations}
\label{eq:Z5res}
\begin{align}
 Z^{\HVBM}_{5}
 &=1+
 \Big(\frac{\alphas}{4\pi}\Big)\Big\{\!-\!4\,C_F\Big\}
 +\Big(\frac{\alphas}{4\pi}\Big)^2\Big\{
  22\,C_F^2
  \!-\!\frac{107}{9}\,C_A
  \!+\!\frac{31}{18}\,C_F N_F
  \Big\}
+\mathcal{O}(\alphas^3)\,.
\label{eq:Z5cdr}
\end{align}
In general, UV renormalized and IR subtracted \FDH\ results differ at most
by terms of $\mathcal{O}(\epsilon^0\,\Neps)$ from the corresponding
quantities in \CDR. Setting $\Neps\!=\!2\epsilon$ and taking the subsequent
limit $\epsilon\!\to\!0$, these differences then vanish. The value of $Z_5^{\HVBM}$
is therefore a regularization-scheme independent quantity to all orders in
perturbation theory,
\begin{align}
 \bar{Z}^{\HVBM}_{5}\equiv Z^{\HVBM}_{5} \, .
 \label{eq:Z5fdh}
\end{align}
\end{subequations}

The regularization-scheme dependent renormalization of operator $O_{\J}^{\HVBM}$
at the one-loop level has first been studied in Ref.\,\cite{Jones:1982zf}.%
\footnote{In this reference, the underlying regularization is called
  'supersymmetric dimensional regularization' (\SDR) which in our nomenclature
  corresponds to dimensional reduction (\DRED). Since, however, contributions
  with external $\epsilon$-scalars vanish for the pseudo-scalar form factors,
  the results coincide with the ones in \FDH.}
For the finite renormalization, the following results are provided,
\begin{align}
 \delta Z_{\text{finite}}^{(1)}=
  \Big(\frac{\alphas}{4\pi}\Big)\Big[\!-\!8\,\frac{C_F}{2}\Big]\,,\qquad
 \delta \bar{Z}_{\text{finite}}^{(1)}=
  \Big(\frac{\alphas}{4\pi}\Big)\Big[\!-\!4\,\frac{C_F}{2}\Big]\,,
  \label{eq:Zfinite}
\end{align}
which are valid in \CDR\ (left) and \FDH\ (right). At first sight,
there seems to be a contradiction to Eq.\,\eqref{eq:Z5fdh}. However,
in Ref.\,\cite{Jones:1982zf} $\Neps$ is identified with $2\epsilon$
throughout the calculation. In this way, contributions from
$\bar{Z}_{\MS}^{\HVBM}$ and $Z_5^{\HVBM}$ are combined. The results in
Eq.\,\eqref{eq:Zfinite} are therefore in agreement with a combination
of Eqs.\,\eqref{eq:ZmsHV} and \eqref{eq:Z5res}.\\

\subsection{Operator renormalization for $\gamma_5^{\text{AC}}$}
\label{sec:acG5uv}

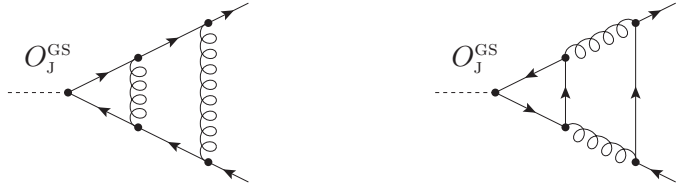
\begin{figure}[t]
\begin{center}
\scalebox{.75}{
\begin{picture}(135,90)(0,10)
\DashLine(0,45)(30,45){2}
\Line[arrow](30,45)(65,62.5)
\Line[arrow](65,62.5)(100,80)
\Line[arrow](100,80)(120,90)
\Line[arrow](120,0)(100,10)
\Line[arrow](100,10)(65,27.5)
\Line[arrow](65,27.5)(30,45)
\Gluon(65,27.5)(65,62.5){4}{4}
\Gluon(100,10)(100,80){4}{8}
\Vertex(30,45){2}
\Vertex(100,80){2}
\Vertex(100,10){2}
\Vertex(65,62.5){2}
\Vertex(65,27.5){2}
\Text(20,62)[c]{\scalebox{1.33}{$O_{\J}^{\GS}$}}
\end{picture}
\begin{picture}(60,90)(0,10)
\Text(30,44)[c]{\scalebox{1.33}{$\phantom{\to}$}}
\end{picture}
\quad
\begin{picture}(135,90)(0,10)
\DashLine(0,45)(30,45){2}
\Line[arrow](30,45)(65,27.5)
\Gluon(65,27.5)(100,10){4}{4}
\Line[arrow](100,10)(100,80)
\Gluon(65,62.5)(100,80){4}{4}
\Line[arrow](65,62.5)(30,45)
\Line[arrow](100,80)(120,90)
\Line[arrow](120,0)(100,10)
\Line[arrow](65,27.5)(65,62.5)
\Vertex(30,45){2}
\Vertex(100,80){2}
\Vertex(100,10){2}
\Vertex(65,62.5){2}
\Vertex(65,27.5){2}
\Text(20,62)[c]{\scalebox{1.33}{$O_{\J}^{\GS}$}}
\end{picture}
}
\end{center}
\caption{\label{fig:fig4}
Sample diagrams contributing to the form factor $\bar{F}_{q,\J}^{\GS}$
at the two-loop level.
}
\end{figure}

In order to determine the so far unknown renormalization of operator
$O_{\J}^{\AC}$, we consider contributions to $\bar{M}_{q,\J}^{\AC}$
up to the two-loop level in the off-shell case. Following
Ref.\,\cite{Adler:1969gk}, it is useful to distinguish two classes of
contributions:
\begin{itemize}
 \item Type A: Contributions where the $\gamma_5^{\AC}$ vertex is
  attached to an external quark line, see the left diagram in Fig.\,\ref{fig:fig4}.
 \item Type B: Contributions where the $\gamma_5^{\AC}$ vertex is
  attached to a quark loop, see the right diagram in Fig.\,\ref{fig:fig4}.
\end{itemize}

\subsubsection*{Type A contributions}
Type A contributions to $\bar{M}_{q,\J}^{\AC}$ can be evaluated in a
particular simple way by applying the setup described in Sec.\,%
\ref{sec:FFg5ca}. Using $(\gamma_{5}^{\AC})^2\!=\!\mathbb{I}$,
all traces can be reduced to expressions without any appearance of
$\gamma_5^{\AC}$. In this way, no difficulties related to the
evaluation of the trace arise. In particular, Type A amplitudes
do not contribute to the anomaly. In analogy to the case of
$\gamma_5^{\HVBM}$, we therefore write the renormalized operator as
\begin{align}
\text{Type A}:\qquad
  \big(O_{\J}^{}\big)_{\text{ren.}}
  \ =\ \bar{Z}_{\MS}^{\AC}\,\big(O_{\J}^{\AC}\big)_{\text{bare}}\,.
  \label{eq:OJACren}
\end{align}
As before, $\bar{Z}_{\MS}^{\AC}$ contains pure poles in $\epsilon$ for
arbitrary $\Neps$. In contrast to Eq.\,\eqref{eq:opRen2}, however, we do
not include a finite renormalization which is due to the fact
that Type A amplitudes are not related to the anomalous contributions to
$\bar{M}_{q,\J}^{\AC}$. The fact that there is no need for the introduction
of symmetry-restoring counterterms at the one-loop level when using
$\gamma_5^{\AC}$ has first been discussed in Ref.\,\cite{Harlander:2005if}.
Further evidence for the validity of Eq.\,\eqref{eq:OJACren} beyond the
one-loop level will be given below.

The so far unknown renormalization constant can be obtained from
an off-shell computation of the amplitudes $\bar{M}_{q,\J}^{\AC}$.
We performed the explicit calculation up to the two-loop level and obtain
the simple result
\begin{align}
\bar{Z}_{\MS}^{\AC}&=1+\mathcal{O}(\alpha_s^3)\,.
\phantom{\Big|}
\label{eq:ZmsACfdh}
\end{align}
The renormalization of operator $O_{\J}^{\AC}$ is therefore trivial, at least
up to two loops. This result is closely related to the use of an anticommutator
in the right definition of Eq.\,\eqref{eq:g5hvneq2a}. If we were to define
$\big[\gamma_{5}^{\AC},\,\gamma^{\mu}_{[\Neps]}\big]\equiv0$ instead,
$\delta\bar{Z}_{\MS}^{\AC}$ would have a non-vanishing value starting at one loop.
In the same way it is the different treatment of strictly $4$- and
$(\dim\!-\!4)$-dimensional quantities in Eqs.\,\eqref{eq:g5hvRules} that
results in the non-vanishing constant $\delta\bar{Z}_{\MS}^{\HVBM}$ given in
Eq.\,\eqref{eq:ZmsHV}.

\subsubsection*{Type B contributions}

\begin{figure}[t]
\begin{center}
\scalebox{.75}{
\begin{picture}(135,90)(0,10)
\DashLine(0,45)(30,45){2}
\Line[arrow](30,45)(65,27.5)
\Gluon(65,27.5)(95,12.5){4}{4}
\DashLine(100,20)(100,70){1.5}
\Gluon(65,62.5)(95,77.5){4}{4}
\Line[arrow](65,62.5)(30,45)
\DashLine(105,82.5)(120,90){1.5}
\DashLine(120,0)(105,7.5){1.5}
\Line[arrow](65,27.5)(65,62.5)
\Vertex(30,45){2}
\Vertex(65,62.5){2}
\Vertex(65,27.5){2}
\Text(20,62)[c]{\scalebox{1.33}{$O_{\J}^{\GS}$}}
\end{picture}
\begin{picture}(60,90)(0,10)
\Text(30,44)[c]{\scalebox{1.33}{$\phantom{\to}$}}
\end{picture}
\quad
\begin{picture}(135,90)(0,10)
\DashLine(0,45)(30,45){2}
\Line[arrow](30,45)(65,27.5)
\DashLine(65,27.5)(95,12.5){4}
\DashLine(100,20)(100,70){1.5}
\DashLine(65,62.5)(95,77.5){4}
\Line[arrow](65,62.5)(30,45)
\DashLine(105,82.5)(120,90){1.5}
\DashLine(120,0)(105,7.5){1.5}
\Line[arrow](65,27.5)(65,62.5)
\Vertex(30,45){2}
\Vertex(65,62.5){2}
\Vertex(65,27.5){2}
\Text(20,62)[c]{\scalebox{1.33}{$O_{\J}^{\GS}$}}
\end{picture}
}
\end{center}
\caption{\label{fig:fig6}
Anomalous (sub)diagrams related to operator $O_{\J}^{\GS}$ with gluons (left)
and $\epsilon$-scalars (right) attached to the loop. The left diagram is only
present in \FDH\ and vanishes according to its Lorentz structure.
}
\end{figure}
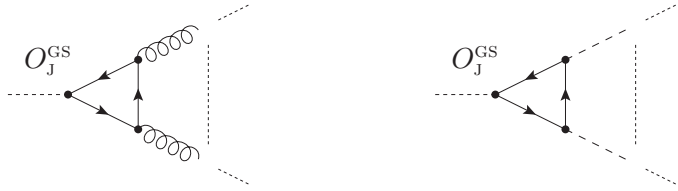

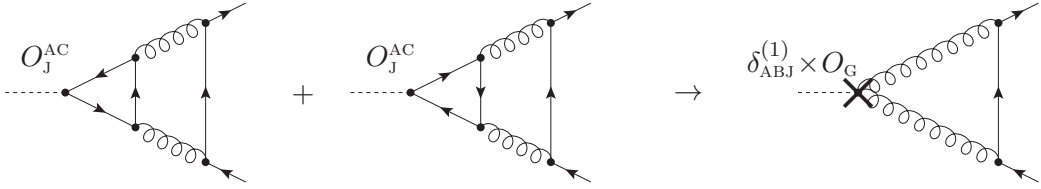
\begin{figure}[t]
\begin{center}
\scalebox{.75}{
\begin{picture}(135,90)(-20,10)
\DashLine(0,45)(30,45){2}
\Line[arrow](30,45)(65,27.5)
\Gluon(65,27.5)(100,10){4}{4}
\Line[arrow](100,10)(100,80)
\Gluon(65,62.5)(100,80){4}{4}
\Line[arrow](65,62.5)(30,45)
\Line[arrow](100,80)(120,90)
\Line[arrow](120,0)(100,10)
\Line[arrow](65,27.5)(65,62.5)
\Vertex(30,45){2}
\Vertex(100,80){2}
\Vertex(100,10){2}
\Vertex(65,62.5){2}
\Vertex(65,27.5){2}
\Text(20,62)[c]{\scalebox{1.33}{$O_{\J}^{\AC}$}}
\end{picture}
\begin{picture}(60,90)(0,10)
\Text(30,44)[c]{\scalebox{1.33}{$+$}}
\end{picture}
\begin{picture}(135,90)(10,10)
\DashLine(0,45)(30,45){2}
\Line[arrow](65,27.5)(30,45)
\Gluon(65,27.5)(100,10){4}{4}
\Line[arrow](100,10)(100,80)
\Gluon(65,62.5)(100,80){4}{4}
\Line[arrow](30,45)(65,62.5)
\Line[arrow](100,80)(120,90)
\Line[arrow](120,0)(100,10)
\Line[arrow](65,62.5)(65,27.5)
\Vertex(30,45){2}
\Vertex(100,80){2}
\Vertex(100,10){2}
\Vertex(65,62.5){2}
\Vertex(65,27.5){2}
\Text(20,62)[c]{\scalebox{1.33}{$O_{\J}^{\AC}$}}
\end{picture}
\begin{picture}(60,90)(10,10)
\Text(30,44)[c]{\scalebox{1.33}{$\to$}}
\end{picture}
\quad
\begin{picture}(135,90)(0,10)
\DashLine(0,45)(30,45){2}
\Gluon(30,45)(100,10){4}{8}
\Line[arrow](100,10)(100,80)
\Gluon(30,45)(100,80){4}{8}
\Line[arrow](100,80)(120,90)
\Line[arrow](120,0)(100,10)
\Vertex(30,45){2}
\Vertex(100,80){2}
\Vertex(100,10){2}
\Text(30,37.5)[b]{\scalebox{2}{\ding{53}}}
\Text(0,62)[c]{\scalebox{1.33}{
  $\delta_{\AB}^{(1)}\!\times\!O_{\G}$}}
\end{picture}
}
\end{center}
\caption{\label{fig5}
Equivalence between anomalous two-loop contributions to $\bar{M}_{q,\J}^{\AC}$
and $\delta_{\AB}^{(1)}\,\bar{M}_{q,\G}^{(1)}$.
}
\end{figure}

Type B contributions 
include traces like in Eq.\,\eqref{eq:actrace}. Let us first consider the
anomalous quark loops shown in Fig.\,\ref{fig:fig6}. These diagrams either
yield direct contributions to the gluon form factor at the one-loop level
or they contribute as subdiagrams at higher loop orders. Their one-loop
result has been obtained in Sec.\,\ref{sec:oneLoopExample2} by using
$\gamma_5^{\HVBM}$ and the \FDF\ framework, respectively. Generalizing to
the case of QCD, we write the corresponding amplitude as
\begin{subequations}
\label{eq:typeBanom}
 \begin{align}
 \big(M_{g,\J}^{\HVBM,(1)}\big)_{\alpha\beta}^{ab}
 &=i\,\Big(\frac{\alphas}{4\,\pi}\Big)\,N_F\,T_F\,
 \delta^{ab}\,\Big\{\epsilon_{\alpha\beta\mu\nu}\Big\}_{[4]}\,
 \Big\{{l_1}^{\mu}\,{l_2}^{\nu}\Big\}_{[d]}
 +\mathcal{O}(\epsilon)\,,
 \label{eq:typeBsubdiag}
 \\
  &\equiv i\, \delta_{\AB}^{(1)}(\alphas)\
 \delta^{ab}\,\Big\{\epsilon_{\alpha\beta\mu\nu}\Big\}_{[4]}\,
 \Big\{{l_1}^{\mu}\,{l_2}^{\nu}\Big\}_{[d]}
 +\mathcal{O}(\epsilon)\,,
 \label{eq:deltaABJ}
 \end{align}
\end{subequations}
where the $l_1$, $l_2$ are line momenta attached to the loop.
Since momenta do not contain evanescent degrees of freedom it
follows that quark loops with external $\epsilon$-scalars vanish.
The fact that the result in Eqs.\,\eqref{eq:typeBanom} is
regularization-scheme independent has first been found in
Ref.\,\cite{Jones:1982zf}.

To obtain a similar result with $\gamma_5^{\AC}$ it is in principle
necessary to modify the trace operation. These redefinitions, however,
are usually made in such a way that they reproduce Eqs.\,\eqref{eq:typeBanom}.
Instead of rederiving the already known result in a different framework
we directly use it in practical computations. This is done by realizing
that the Lorentz and the color structure in Eqs.\,\eqref{eq:typeBanom} are
exactly the same as in the Feynman rule given in Eq.\,\eqref{frO3a}.
Accordingly, for Type B contributions the renormalization of operator
$O_{\J}$ is closely related to the one of $O_{\G}$, see Fig.\,\ref{fig5}.
Up to the two-loop level we therefore write
\begin{align}
 \text{Type B}:\qquad
 \big(O_{\J}^{\AC}\big)_{\text{ren.}}\ &\equiv\
 \delta_{\AB}^{(1)}(\alphas)\times\big(O_{\G}\big)_{\text{ren.}}
  +\mathcal{O}(\alphas^3)\,.
  \label{eq:TypeBren}
\end{align}
In this way, $\gamma_5$ is effectively removed from the computation.
The necessary one-loop renormalization of operator $O_{\G}$ does not depend
on the treatment of $\gamma_5$ and is known from Sec.\,\ref{sec:bmG5uv}.%
\footnote{The approach of evaluating Type A contributions using $\gamma_5^{\AC}$
  and Type B contributions using $\gamma_5^{\HVBM}$ has been discussed before in
  Ref.\,\cite{Bernreuther:2004ih}. In this reference, however, the right diagram
  in   Fig.\,\ref{fig5} is evaluated as a whole by using projections that
  lead to similar expressions as in Eq.\,\eqref{eq:epsSqD}. Accordingly,
  the $\varepsilon$ pseudotensor is treated in $\dim\!\neq\!4$
  dimensions and additional finite counterterms have to be added to
  obtain the correct result.
  In Eq.\,\eqref{eq:TypeBren}, on the other hand, the known $\mathcal{O}(\alphas)$
  value of the anomaly is used to effectively reduce the evaluation of 
  the two-loop diagram to a one-loop problem that does not depend on the 
  specific treatment of $\gamma_5$.}

\subsubsection*{Comparison of BM and AC}

With the results of the previous sections it is possible to compare the
UV-renormalized off-shell values of $\bar{\mathcal{F}}_{q,\J}^{\GS}$
obtained in \HVBM\ and \AC,
\begin{subequations}
\label{eq:Z5det}
\begin{align}
 \bar{\mathcal{F}}_{q,\J}^{\HVBM}
  &=Z_{5}^{\HVBM}\,\bar{Z}_{\MS}^{\HVBM}\,
  \big(\bar{F}_{q,\J}^{\HVBM}\big)_{\text{ren.}}
  \ +\ \mathcal{O}(\alphas^3)\,,
  \\*[.2cm]
  \bar{\mathcal{F}}_{q,\J}^{\AC}
  &=\underbrace{
    \big(\bar{F}_{q,\J}^{\AC}\big)_{\text{ren.}}
    \phantom{\Big|}
    }_{\text{Type A}}
  +\ \underbrace{
    \delta_{\AB}^{(1)}\,\Big[
      \bar{R}_{q,\G}^{(1)}/R_{q,\J}^{(0),\AC}
      +\delta\bar{Z}_{\G\J}^{(1)}
      \Big]
      }_{\text{Type B}}
   \ +\ \mathcal{O}(\alphas^3)\,.
\end{align}
\end{subequations}
The subscript 'ren.' indicates that a coupling, gauge parameter, and field
(sub)renormalization is applied to the bare coefficients. Taking the limit
$\epsilon\!\to\!0$, we find that both results in Eqs.\,\eqref{eq:Z5det}
coincide,
\begin{align}
\text{offshell:}\qquad
  \bar{\mathcal{F}}_{q,\J}^{\HVBM}\,\Big|_{\epsilon\to 0}
  \ \equiv\
  \bar{\mathcal{F}}_{q,\J}^{\AC}\,\Big|_{\epsilon\to 0}+\mathcal{O}(\alpha_s^3)\,.
  \label{eq:FFrenEq}
\end{align}
This provides further evidence for the fact that there is no need for the
introduction of finite counterterms when using $\gamma_5^{\AC}$.
Compared to $\HVBM$, therefore not only the evaluation of the algebra is
much simpler but also the renormalization of operator $O_{\J}^{\AC}$.

Extending these considerations to higher loop orders, it is possible
to determine the so far unknown three-loop value of $Z_5^{\HVBM}$ from
a genuine three-loop calculation.
So far, the standard way to obtain $Z_5^{\HVBM}$ is to consider the
(anomalous) relation between the axial-vector and the pseudo-scalar
current in the effective theory
for the case of $N_F$ massless quarks
and to evaluate it between two gluon states (see e.\,g.\ Ref.\,\cite{Larin:1993tq}).
Since the anomaly itself is of $\mathcal{O}(\alphas)$, however, the $l$-loop
coefficient of $Z_5^{\HVBM}$ has to be obtained from an $(l\!+\!1)$-loop
calculation. In contrast, using an extension of Eqs.\,\eqref{eq:Z5det} and
\eqref{eq:FFrenEq} beyond the two-loop level allows one to determine the same
coefficient from an $l$-loop calculation.

\subsection{UV renormalized form factors}
\label{sec:uvRenFF}

Using the results of the renormalization constants from the previous sections
together with Eqs.\,\eqref{eq:FFren}, the UV renormalized form factors in the
\FDH\ scheme finally read
\begin{subequations}
\small
\label{eq:uvRes}
\begin{align}
\bar{\mathcal{F}}_{q,\J}^{\HVBM}
&=
1\!+\!
\Big(\frac{\alphas}{4\pi}\Big) \bigg\{
  C_F \Big[
    \!-\!\frac{2}{\epsilon^2}
    \!-\!\frac{3}{\epsilon}
    \!-\!5
    \!+\!\frac{\pi^2}{6}
    \!+\!\epsilon\Big(
      3
      \!+\!\frac{\pi^2}{4}
      \!+\!\frac{14}{3}\zeta(3)
      \Big)
    \!-\!\epsilon^2\Big(
      3
      \!-\!\frac{\pi^2}{4}
      \!-7\zeta(3)
      \!-\!\frac{47}{720}\pi^4
      \Big)
    \Big]
   \!+\!\mathcal{O}(\epsilon^3)
   \bigg\}
\notag\\&\quad
+\Big(\frac{\alphas}{4\pi}\Big)^2 \bigg\{
  C_A C_F \Big[
    \frac{11}{2\epsilon^3}
    \!+\!\frac{
      \frac{23}{18}
      \!+\!\frac{\pi^2}{6}
      }{\epsilon^2}
    \!-\!\frac{
      \frac{1075}{108}
      \!+\!\frac{11}{12}\pi^2
      \!-\!13\zeta(3)
      }{\epsilon}
    \!-\!\frac{25279}{648}
    \!-\!\frac{46}{27}\pi^2
    \!+\!\frac{313}{9}\zeta(3)
    \!+\!\frac{11}{45}\pi^4
    \Big]
\notag\\&\quad
  +C_F^2 \Big[
    \frac{2}{\epsilon^4}
    \!+\!\frac{6}{\epsilon^3}
    \!+\!\frac{
      \frac{29}{2}
      \!-\!\frac{\pi^2}{3}
      }{\epsilon^2}
    \!+\!\frac{
      \frac{77}{4}
      \!-\!\frac{64}{3}\zeta (3)
      }{\epsilon}
    \!+\!\frac{139}{8}
    \!-\!\frac{\pi^2}{4}
    \!-\!58\,\zeta (3)
    -\!\frac{13}{36}\pi^4
    \Big]
\notag\\&\quad
  +C_F N_F \Big[
    \!-\!\frac{1}{\epsilon^3}
    \!-\!\frac{4}{9\epsilon^2}
    \!+\!\frac{
      \frac{46}{27}
      \!+\!\frac{\pi^2}{6}
      }{\epsilon}
      \!-\!\frac{1679}{162}
      \!+\!\frac{23}{54}\pi^2
      \!+\!\frac{2}{9}\zeta(3)
      \Big]
  +\mathcal{O}(\epsilon)
  \bigg\}
  +\mathcal{O}(\alphas^3)\,,
\phantom{\Big|}
\\[.35cm]
\bar{\mathcal{F}}_{q,\G}&=
\Big(\frac{\alphas}{4\pi}\Big) \bigg\{
  C_A \Big[
    \frac{7115}{324}
    \!-\!\frac{\pi^2}{9}
    \!-\!2\,\zeta (3)
    +\epsilon\Big(
      \frac{111049}{1944}
      \!-\!\frac{7321}{11664}\pi^2
      \!-\!8\,\zeta(3)
      \!-\!\frac{53}{1620}\pi^4
      \!-\!\frac{\pi^2}{18}\zeta(3)
      \Big)
\notag\\&\quad
    \!+\!\epsilon^2\Big(
      \frac{660451}{3888}
      \!-\!\frac{17335}{7776}\pi^2
      \!-\!\frac{80515}{2916}\zeta(3)
      \!-\!\frac{300449}{2099520}\pi^4
      \!-\!20\,\zeta(5)
      \!-\!\frac{53}{58320}\pi^6
      \!-\!\frac{19}{81}\pi^2\zeta (3)
      \!-\!\frac{14}{9}\zeta(3)^2
\notag\\&\quad
      \!-\!\frac{\pi^4}{648}\zeta(3)
      \Big)
    \Big]
   +C_F \Big[
    \!-\!\frac{2}{\epsilon^2}
    \!-\!\frac{3}{\epsilon}
    \!-\!\frac{29}{4}
    \!+\!\frac{\pi^2}{6}
    \!+\!\epsilon\,\Big(
      \!-\!\frac{203}{24}
      \!-\!\frac{\pi^2}{16}
      \!+\!\frac{14}{3}\zeta(3)
      \Big)
\notag\\&\quad
    \!+\!\epsilon^2\,\Big(
      \!-\!\frac{1115}{144}
      \!-\!\frac{947}{864}\pi^2
      \!+\!\frac{127}{12}\zeta(3)
      \!+\!\frac{163}{2880}\pi^4
      \Big)
    \Big]
    +N_F \Big[
      \!-\!\frac{445}{162}
      \!+\!\epsilon\,\Big(
	\!-\!\frac{8231}{972}
	\!+\!\frac{239}{5832}\pi^2
	\!+\!\frac{4}{3}\zeta(3)
	\Big)
\notag\\&\quad
      \!+\!\epsilon^2\,\Big(
	\!-\!\frac{50533}{1944}
	\!+\!\frac{1835}{11664}\pi^2
	\!+\!\frac{9125}{1458}\zeta(3)
	\!+\!\frac{22903}{1049760}\pi^4
	\!+\!\frac{1}{27}\pi^2\zeta (3)
	\Big)
      \Big]
      +\mathcal{O}(\epsilon^3)
   \bigg\}
  +\mathcal{O}(\alphas^2)\,,
  \phantom{\Big|}
\\
\bar{\mathcal{F}}_{g,\G}&=1\!+\!
 \Big(\frac{\alphas}{4\pi}\Big) \bigg\{
  C_A \Big[
    \!-\!\frac{2}{\epsilon^2}
    \!-\!\frac{11}{3\epsilon}
    \!+\!\frac{13}{3}
    \!+\!\frac{\pi^2}{6}
    \!+\!\epsilon\Big(
      12
      \!+\!\frac{14}{3}\zeta (3)
      \Big)
    \!+\!\epsilon^2\Big(
      28
      \!-\!\frac{\pi^2}{3}
      \!+\!\frac{47\pi^4}{720}
      \Big)
    \Big]
   \!+\!\frac{2 N_F}{3\epsilon}
 +\mathcal{O}(\epsilon^3)
 \bigg\}
\notag\\&\quad
 +\Big(\frac{\alphas}{4\pi}\Big)^2 \bigg\{
  C_A^2 \Big[
    \frac{2}{\epsilon^4}
    \!+\!\frac{77}{6\epsilon^3}
    \!+\!\frac{
      \frac{5}{9}
      \!-\!\frac{\pi^2}{6}
      }{\epsilon^2}
    \!-\!\frac{
      \frac{1444}{27}
      \!+\!\frac{11}{36}\pi^2
      \!+\!\frac{25}{3}\zeta(3)
      }{\epsilon}
    \!-\!\frac{2882}{81}
    \!+\!\frac{29}{9}\pi^2
    \!-\!33\zeta (3)
    \!-\!\frac{7}{60}\pi^4
    \Big]
\notag\\&\quad
  +C_A N_F \Big[
    \!-\!\frac{7}{3\epsilon^3}
    \!-\!\frac{13}{3\epsilon^2}
    \!+\!\frac{
      \frac{148}{27}
      \!+\!\frac{\pi^2}{18}
      }{\epsilon}
    \!-\!\frac{295}{81}
    \!-\!\frac{5}{18}\pi^2
    \!-\!2\zeta(3)
    \Big]
\notag\\&\quad
  +C_F N_F \Big[
    \frac{1}{\epsilon}
    \!-\!\frac{74}{3}
    \!+\!8\zeta(3)
    \Big]
  +N_F^2\,\frac{4}{9\,\epsilon^2}
  +\mathcal{O}(\epsilon)
  \bigg\}
+\mathcal{O}(\alphas^3)\, ,
\\[.35cm]
\bar{\mathcal{F}}_{g,\J}^{\HVBM}&=  
  \Big(\frac{\alphas}{4\pi}\Big) \bigg\{
    C_A \Big[
      \!-\!\frac{2}{\epsilon^2}
      \!-\!\frac{11}{3\epsilon}
      \!+\!\frac{13}{3}
      \!+\!\frac{\pi^2}{6}
      \!+\!\epsilon\Big(
	16
	\!-\!\frac{\pi^2}{3}
	\!+\!\frac{32}{3}\zeta(3)
	\Big)
      \!+\!\epsilon^2\Big(
	\frac{152}{3}
	\!-\!\frac{4}{3}\pi^2
	\!+\!2\zeta(3)
	\!+\!\frac{127}{720}\pi^4
	\Big)
      \Big]
\notag\\*&\quad
    +C_F \Big[
      \epsilon \Big(
	10
	\!-\!12\,\zeta (3)
	\Big)
      +\epsilon^2\,\Big(
	38
	\!-\!\frac{7}{6}\pi^2
	\!-\!18\,\zeta(3)
	\!-\!\frac{\pi^4}{5}
      \Big)
    \Big]
    \!+\!\frac{2N_F}{3\,\epsilon}
    \!+\!\mathcal{O}(\epsilon^3)
  \bigg\}
  +\mathcal{O}(\alphas^3)\, .
  \phantom{\Big|}
\end{align}
\end{subequations}
\normalsize
Compared to the \CDR\ results which are given e.\,g.\ in Ref.\,\cite{Ahmed:2015qpa},
the one-loop coefficients differ by terms of $\mathcal{O}(\epsilon^0)$, whereas at
the two-loop level these differences are of $\mathcal{O}(\epsilon^{-2})$. After
subtracting the IR divergences and taking the physical limit $\epsilon\!\to\!0$,
however, we obtain the same (regularization-scheme independent) results.

\section{Conclusions}
\label{sec:conc}

In this article we discussed the regularization-scheme dependent treatment
of $\gamma_5$ within dimensional regularization. So far, \CDR\ in combination
with $\gamma_5^{\HVBM}$ as defined in Eq.\,\eqref{eq:g5hvbm} has been the
most commonly used approach to perform perturbative computations in the
dimensional framework.
One main reason might be that the approach is based on an explicit construction
prescription which enables the use of standard calculational techniques
like cyclicity of the trace. At the practical level, however, the evaluation
of the algebra is cumbersome due to the increased number of $\gamma$ matrices
and the ad hoc (anti)symmetrization of $\gamma_5^{\HVBM}$ operators.
Moreover, since initial symmetries are broken explicitly there is an immanent
need for the introduction of additional counterterms to obtain correct results.
In comparison, the application of an anticommuting $\gamma_5^{\AC}$
simplifies the evaluation of the Lorentz algebra significantly which is due
to the fact that algebraic properties remain unchanged compared to the
unregularized theory. This, however, is not the case for $\gamma_5^{\AC}$-odd
traces. Since these either vanish or do not exhibit cyclicity, special attention
has to be paid to the non-breaking of gauge invariance and other symmetries of
the underlying theory.

At the one-loop level, the \FDF\ approach avoids all these complications related
to the treatment of $\gamma_5$ since, using a strictly four-dimensional algebra,
the matrices $\gamma_5^{\HVBM}$ and $\gamma_5^{\AC}$ as well as their algebraic
behavior are identical.
In Secs.\,\ref{sec:oneLoopExample1} and \ref{sec:oneLoopExample2}, \FDF\ has
proven as an effective implementation of the Lorentz algebra that reduces the
technical complexity significantly, even including contributions to the axial
anomaly. At the same time the results are compatible with gauge
invariance and Bose symmetry. In the examples considered, the \FDF\ results
are entirely given by so-called extra integrals which can be evaluated in a
particular simple way. The question whether \FDF\ can be extended beyond the
one-loop level, such that it leads to a facilitation compared to more
traditional schemes remains to be answered.

At the two-loop level, we investigated the possibility of utilizing the
benefits of different $\gamma_5$ schemes and computed the pseudo-scalar
form factors of quarks and gluons in the \FDH\ scheme. We have shown
explicitly that evanescent Higgs--$\epsilon$-scalar interactions are absent
and determined the so far unknown UV renormalization of the corresponding
operators. The results of the UV-renormalized form factors are compatible
with the general prediction for IR divergences in \FDH.
As a general recommendation for the treatment of $\gamma_5$ we find that the
use of $\gamma_5^{\HVBM}$ should be avoided whenever $\gamma_5^{\AC}$ leads
to an obvious and immediate simplification. This clearly applies to Type~A
contributions to the pseudo-scalar form factors where not only the evaluation
of analytical expressions is simplified but also the operator renormalization
when using $\gamma_5^{\AC}$. It should be mentioned explicitly that these
simplifications are not restricted to \FDH\ but apply to all considered
dimensional schemes.
For the evaluation of (anomalous) $\gamma_5$-odd expressions (like Type~B
contributions in Sec.\,\ref{sec:acG5uv}), however, it is not clear at all,
if the use of $\gamma_5^{\AC}$ leads to a perceptible simplification due to
the aforementioned complications. In this case, the use of $\gamma_5^{\HVBM}$
therefore still constitutes a viable alternative. Moreover, seizing a
suggestion of Ref.\,\cite{Jegerlehner:2000dz}, Type~B contributions can be
obtained by removing $\gamma_5$ in analytical expressions altogether.
This can be done by using the well-known and scheme-independent results of
the anomalies. In this way not only the evaluation of the amplitudes is
significantly simplified but also the related operator renormalization.

Finally it should be mentioned that observables related to Lagrangian~%
\eqref{eq:Lfull} are usually obtained in an effective theory for the case
of massless quarks. One requirement for this option is that the Lorentz
structure related to the $\gamma_5$ vertex can be effectively disentangled
from the kinematics of the underlying process.
As it turned out, for the schemes considered in this article this is
only possible for $\gamma_5^{\HVBM}$. For the choice of a particular
$\gamma_5$ scheme one therefore has to compare the complexity of a
calculation with massless quarks and the extended $\gamma_5^{\HVBM}$
algebra with the complexity of a calculation with massive quarks and
a simplified $\gamma_5^{\AC}$ algebra. The decision which of these
alternatives is the more efficient one remains to be made on an
individual basis.

\bigskip

\bigskip

\appendix
\newpage
\section{Appendix}
\subsection{Feynman rules and definition of the form factors}
\label{sec:FeynmanRules}

The Feynman rules originating from the effective Lagrangian~\eqref{eq:Leff} read
\begin{subequations}
\label{eq:FeynmanRules}
\begin{align}
  \scalebox{.5}{
 \begin{picture}(110,0)(20,50)
 \Vertex(35,50){2}
 \DashLine(0,50)(35,50){2}
 \ArrowLine(35,50)(100,100)
 \ArrowLine(35,50)(100,0)
 \LongArrow(80,96)(60,80)
 \LongArrow(80,6)(60,21)
 \Text(63, 100)[c]{\scalebox{2}{$l_2$}}
 \Text(63, 5  )[c]{\scalebox{2}{$l_1$}}
 \Text( 15, 70)[c]{\scalebox{2}{$O_{\J}^{\HVBM}$}}
 \Text(115,105)[c]{\scalebox{2}{$\overline{\psi}_{i}$}}
 \Text(115,-5)[c]{\scalebox{2}{$\psi_{j}$}}
 \label{frO1a}
 \end{picture} }
 &=\ \frac{-i}{4!\,3!\,2}\,\delta_{ij}\,
 \Big\{\epsilon_{\mu\nu\rho\sigma}\Big\}_{[4]}\,
 \Big\{
  (l_{1}+l_{2})^{\mu}\,\gamma^{\nu}\gamma^{\rho}\gamma^{\sigma}
  \pm\text{perm.}
  \Big\}_{[\dim]},
 \phantom{\begin{aligned}|\\ |\\ |\end{aligned}}
  \\[.75cm]
  \scalebox{.5}{
 \begin{picture}(110,0)(20,50)
 \Vertex(35,50){2}
 \DashLine(0,50)(35,50){2}
 \ArrowLine(35,50)(100,100)
 \ArrowLine(35,50)(100,0)
 \LongArrow(80,96)(60,80)
 \LongArrow(80,6)(60,21)
 \Text(63, 100)[c]{\scalebox{2}{$l_2$}}
 \Text(63, 5  )[c]{\scalebox{2}{$l_1$}}
 \Text( 15, 70)[c]{\scalebox{2}{$O_{\J}^{\AC}$}}
 \Text(115,105)[c]{\scalebox{2}{$\overline{\psi}_{i}$}}
 \Text(115,-5)[c]{\scalebox{2}{$\psi_{j}$}}
 \label{frO2a}
 \end{picture} }
 &=\ -i\,\delta_{ij}\,
 (\slashed{l}_{1}+\slashed{l}_{2})\,\gamma_5^{\AC}\,,
 \phantom{\begin{aligned}|\\ |\\ |\end{aligned}}
 \\[.75cm]
 \scalebox{.5}{
 \begin{picture}(110,0)(20,50)
 \Vertex(35,50){2}
 \DashLine(0,50)(35,50){2}
 \Gluon(35,50)(100,100){5}{6}
 \Gluon(35,50)(100,0){5}{6}
 \LongArrow(80,99)(60,83)
 \LongArrow(80,3)(60,18)
 \Text(63, 100)[c]{\scalebox{2}{$l_2$}}
 \Text(63, 0  )[c]{\scalebox{2}{$l_1$}}
 \Text(15, 70)[c]{\scalebox{2}{$O_{\G}$}}
 \Text(115, -5)[c]{\scalebox{2}{$A_{\alpha}^{a}$}}
 \Text(115,105)[c]{\scalebox{2}{$A_{\beta }^{b}$}}
 \label{frO3a}
 \end{picture} }
 &=\ -\frac{i}{4!}\,\delta^{ab}\,\Big\{\epsilon_{\mu\nu\rho\sigma}\Big\}_{[4]}\,
 \Big\{
  l_{1}^{\mu}\,l_{2}^{\nu}\,
  g_{\phantom{\rho}\alpha}^{\rho}\,g_{\phantom{\sigma}\beta}^{\sigma}
  \pm\text{perm.}
  \Big\}_{[\dim]},
 \phantom{\begin{aligned}|\\ |\\ |\end{aligned}}
 \\[.75cm]
 \scalebox{.5}{
 \begin{picture}(110,0)(20,50)
 \Vertex(35,50){2}
 \DashLine(0,50)(35,50){2}
 \Gluon(35,50)(100,100){5}{6}
 \Gluon(35,50)(100,50){5}{6}
 \Gluon(35,50)(100,0){5}{6}
 \LongArrow(80,99)(60,83)
 \LongArrow(80,3)(60,18)
 \LongArrow(82,60)(62,60)
 \Text( 63,100)[c]{\scalebox{2}{$l_3$}}
 \Text( 95, 65)[c]{\scalebox{2}{$l_2$}}
 \Text( 63,  0)[c]{\scalebox{2}{$l_1$}}
 \Text( 15, 70)[c]{\scalebox{2}{$O_{\G}$}}
 \Text(115, -5)[c]{\scalebox{2}{$A_{\alpha}^{a}$}}
 \Text(115, 50)[c]{\scalebox{2}{$A_{\beta }^{b}$}}
 \Text(115,105)[c]{\scalebox{2}{$A_{\gamma}^{c}$}}
 \label{frO4a}
 \end{picture} }
 &=\ -\frac{i}{4!}\,f^{abc}\,\Big\{\epsilon_{\mu\nu\rho\sigma}\Big\}_{[4]}\,
  \Big\{
  (l_{1}+l_{2}+l_{3})^{\mu}\,
  g_{\phantom{\nu}\alpha}^{\nu}\,g_{\phantom{\rho}\beta}^{\rho}\,
  g_{\phantom{\sigma}\gamma}^{\sigma}
  \pm\text{perm.}
  \Big\}_{[\dim]},
 \phantom{\begin{aligned}|\\ |\\ |\end{aligned}}
 \\ \nonumber
\end{align}
\end{subequations}
where 'perm.' denotes terms originating from further permutations in the
indices $\mu,\nu,\rho,\sigma$.

According to the discussion in Sec.\,\ref{sec:epstensor}, we decompose
pseudo-scalar amplitudes as
\begin{subequations}
 \begin{align}
  \bar{M}_{q,\J}^{\HVBM}
  &=\Big\{\epsilon_{\mu\nu\rho\sigma}\Big\}_{[4]}\ \bar{u}(p_1)\,\sum_{n=0}
    \Big\{
    \big(\bar{R}_{q,\J}^{\HVBM,(n)}\big)^{\mu\nu\rho\sigma}\Big\}_{[d]}\,
    v(p_2)\,,
  \\
  \bar{M}_{q,\G}
  &=\Big\{\epsilon_{\mu\nu\rho\sigma}\Big\}_{[4]}\ \bar{u}(p_1)\,\sum_{n=1}
    \Big\{
    \big(\bar{R}_{q,\G}^{(n)}\big)^{\mu\nu\rho\sigma}\Big\}_{[d]}\,
    v(p_2)\,,
  \\
  \bar{M}_{g,\G}
  &=\Big\{\epsilon_{\mu\nu\rho\sigma}\Big\}_{[4]}\ \sum_{n=0}
    \Big\{\big(
    \bar{R}_{g,\G}^{(n)}\big)^{\mu\nu\rho\sigma}_{\alpha\beta}\Big\}_{[d]}\,
    \epsilon^{\alpha}(p_1)\,\epsilon^{\beta}(p_2)\,,
  \\
  \bar{M}_{g,\J}^{\HVBM}
  &=\Big\{\epsilon_{\mu\nu\rho\sigma}\Big\}_{[4]}\ \sum_{n=1}
    \Big\{\big(
    \bar{R}_{g,\J}^{\HVBM,(n)}\big)^{\mu\nu\rho\sigma}_{\alpha\beta}\Big\}_{[d]}\,
    \epsilon^{\alpha}(p_1)\,\epsilon^{\beta}(p_2)\,,
 \end{align}
\end{subequations}
where $v,u$ are (anti)quark spinors and $\epsilon^{\mu}$ are polarization
vectors of the gluon. The sum of the (outgoing) momenta $p_1$ and $p_2$ is given by
$p_1\!+\!p_2\!=\!q$.
According their Lorentz decomposition, the remainders can
be written as
\begin{subequations}
\label{eq:remainders}
 \begin{align}
  \big(\bar{R}_{q,\J}^{\HVBM,(n)}\big)^{\mu\nu\rho\sigma}
  &\equiv\phantom{:}\bar{R}_{q,\J}^{\HVBM,(n)}\,
    \Big\{
      q^{\mu}\gamma^{\nu}\gamma^{\rho}\gamma^{\sigma}
      \pm \text{perm.}\Big\}
  \equiv\bar{R}_{q,\J}^{\HVBM,(n)}\,
    \big(P_{q}^{}\big)^{\mu\nu\rho\sigma}\,,\phantom{\Big|}
  \\*
  \big(\bar{R}_{q,\G}^{(n)}\big)^{\mu\nu\rho\sigma}
  &\equiv\phantom{:}
  \bar{R}_{q,\G}^{(n)}\,
    \big(P_{q}^{}\big)^{\mu\nu\rho\sigma}\,,\phantom{\Big|}
  \\*
  \big(\bar{R}_{g,\G}^{(n)}\big)^{\mu\nu\rho\sigma}_{\alpha\beta}
  &\equiv\phantom{:}
    \bar{R}_{g,\G}^{(n)}\,\phantom{\Big|}
    \Big\{
      p_1^{\mu}\,p_2^{\nu}\,
	g^{\rho}_{\phantom{\rho}\alpha}g^{\sigma}_{\phantom{\sigma}\beta}
            \pm \text{perm.}\Big\}
   \equiv \bar{R}_{g,\G}^{(n)}\,
      \big(P_{g}\big)^{\mu\nu\rho\sigma}_{\alpha\beta}\,,
  \\*
  \big(\bar{R}_{g,\J}^{\HVBM,(n)}\big)^{\mu\nu\rho\sigma}_{\alpha\beta}
    &\equiv\phantom{:}
   \bar{R}_{g,\J}^{\HVBM,(n)}\,
    \big(P_{g}\big)^{\mu\nu\rho\sigma}_{\alpha\beta}\,,\phantom{\Big|}
 \end{align}
\end{subequations}
For the extraction of the coefficients on the r.\,h.\,s.\ of
Eqs.\,\eqref{eq:remainders} we define the following normalization factors,
\begin{subequations}
  \begin{align}
   \text{Tr}\Big[
    q_{\mu}\,\gamma_{\nu}\gamma_{\rho}\gamma_{\sigma}\,
    \big(P_{q}^{}\big)^{\mu\nu\rho\sigma}
    \Big]
    &=\frac{1}{6}(d\!-\!1)\,(d\!-\!2)\,(d\!-\!3)\,q^2
    \equiv N_{q}^{}\, ,
   \\*[.25cm]
   \big(P_{g}\big)^2
    &=\!-\frac{1}{144}(d\!-\!2)\,(d\!-\!3)\,q^4
    \equiv N_{g}\, .
  \end{align}
\end{subequations}
The remainders entering Eqs.\,\eqref{eq:bareFF} are then obtained by
\begin{subequations}
 \begin{align}
  \bar{R}_{q,\J}^{\HVBM,(n)}
  &=\big(N_{q}^{}\big)^{-1}\,\text{Tr}\Big[
    q_{\mu}\,\gamma_{\nu}\gamma_{\rho}\gamma_{\sigma}\
    \big(\bar{R}_{q,\J}^{\HVBM,(n)}\big)^{\mu\nu\rho\sigma}
    \Big]\, ,
  \\*[.25cm]
  \bar{R}_{q,\G}^{(n)}
  &=\big(N_{q}^{}\big)^{-1}\,\text{Tr}\Big[
    q_{\mu}\,\gamma_{\nu}\gamma_{\rho}\gamma_{\sigma}\
    \big(\bar{R}_{q,\G}^{(n)}\big)^{\mu\nu\rho\sigma}
    \Big]\, ,
  \\*[.25cm]
  \bar{R}_{g,\G}^{(n)}
  &=\big(N_{g}\big)^{-1}\,
    \big(P_{g}\big)_{\mu\nu\rho\sigma}^{\alpha\beta}\
      \big(\bar{R}_{g,\G}^{(n)}\big)^{\mu\nu\rho\sigma}_{\alpha\beta}\, ,
  \\*[.25cm]
  \bar{R}_{g,\J}^{\HVBM,(n)}
  &=\big(N_{g}\big)^{-1}\,
    \big(P_{g}\big)_{\mu\nu\rho\sigma}^{\alpha\beta}\
      \big(\bar{R}_{g,\J}^{\HVBM,(n)}\big)^{\mu\nu\rho\sigma}_{\alpha\beta}\,.
 \end{align}
\end{subequations}

\subsection{Bare on-shell results}
\label{sec:bareFF}

The non-vanishing coefficients of the bare form factors defined in
Eqs.~\eqref{eq:bareFF} read
\small
\begin{subequations}
\begin{align}
\bar{F}_{q,\J}^{\HVBM,(1)}&=
  \Big(\frac{\alphas}{4\pi}\Big)C_F\Big[
    \!-\!\frac{2}{\epsilon^2}
    \!-\!\frac{3}{\epsilon}
    \!-\!2
+\!\frac{\pi^2}{6}
    \!+\!\epsilon\Big(
      2
      \!+\!\frac{\pi^2}{4}
      \!+\!\frac{14}{3}\zeta(3)
      \Big)
    \!+\!\epsilon^2\Big(
      10
      \!+\!\frac{\pi^2}{6}
      \!+\!7\zeta (3)
      \!+\!\frac{47}{720}\pi^4
      \Big)
    \Big]
\notag\\&\quad
    +\Big(\frac{\alphae}{4\pi}\Big)\,C_F\,\Big[
    \!-\!1
    \!-\!5\epsilon
    \!+\!\epsilon^2\Big(
      \!-\!13
      \!+\!\frac{\pi^2}{12}
      \Big)
    \Big]+\mathcal{O}(\epsilon^3)\,,
\label{eq:FgJ1bare}
\\
\bar{F}_{q,\J}^{\HVBM,(2)}&=
 \Big(\frac{\alphas}{4\pi}\Big)^2 \bigg\{
  C_A C_F \Big[
    \!-\!\frac{11}{6\epsilon^3}
    \!-\!\frac{
      \frac{163}{18}
      \!-\!\frac{\pi^2}{6}
      }{\epsilon^2}
    -\frac{
      \frac{2551}{108}
      +\frac{11}{36}\pi^2
      \!-\!13\,\zeta (3)
      }{\epsilon}
    \!-\!\frac{23623}{648}
    \!-\!\frac{91}{108}\pi^2
    \!+\!\frac{467}{9}\zeta (3)
    \!+\!\frac{11}{45}\pi^4
    \bigg]
\notag\\&\quad
  \!+\!C_F^2\Big[
    \frac{2}{\epsilon^4}
    \!+\!\frac{6}{\epsilon^3}
    \!+\!\frac{
      \frac{21}{2}
      \!-\!\frac{\pi^2}{3}
      }{\epsilon^2}
    \!+\!\frac{
      \frac{53}{4}
      \!-\!\frac{64}{3}\zeta(3)
      }{\epsilon}
    \!-\!\frac{53}{8}
    \!+\!\frac{\pi^2}{12}
    \!-\!58\zeta (3)
    \!-\!\frac{13}{36}\pi^4
    \Big]
\notag \\&\quad
  \!+C_F N_F \bigg[
    \frac{1}{3\epsilon^3}
    \!+\!\frac{14}{9\epsilon^2}
    \!+\!\frac{
      \frac{37}{27}
      \!+\!\frac{\pi^2}{18}
      }{\epsilon}
    \!-\!\frac{1283}{162}
    \!+\!\frac{7}{27}\pi^2
    \!-\!\frac{26}{9}\zeta(3)
    \bigg]
  \bigg\}
  \!+\mathcal{O}(\epsilon^1) \, ,
\\[.4cm]
\bar{F}_{q,\G}^{(1)}&=
  \Big(\frac{\alphas}{4\pi}\Big) \bigg\{
    C_A \Big[
      \frac{11}{3\,\epsilon }
      \!+\!\frac{263}{18}
      \!+\!\epsilon\,\Big(
	\frac{4949}{108}
	\!-\!\frac{23}{36}\pi^2
	\!-\!6\,\zeta(3)
	\Big)
      \!+\!\epsilon^2\,\Big(
	\frac{87917}{648}
	\!-\!\frac{479}{216}\pi^2
	\!-\!\frac{257}{9}\zeta(3)
	\!-\!\frac{4}{45}\pi^4
	\Big)
      \Big]
\notag\\&\quad
    \!+C_F \Big[
      \!-\!\frac{2}{\epsilon^2}
      \!-\!\frac{3}{\epsilon}
      \!-\!\frac{11}{2}
      \!+\!\frac{\pi^2}{6}
      \!+\!\epsilon\Big(
	\!-\!\frac{37}{4}
	\!+\!\frac{\pi^2}{4}
	\!+\!\frac{14}{3}\zeta(3)
	\Big)
      \!+\!\epsilon^2\Big(
	\!-\!\frac{103}{8}
	\!-\!\frac{13}{24}\pi^2
	\!+\!7\zeta(3)
	\!+\!\frac{47}{720}\pi^4
	\Big)
      \Big]
\notag\\&\quad
    \!+\!N_F \Big[
      \!-\!\frac{2}{3\,\epsilon}
      \!-\!\frac{19}{9}
      \!-\epsilon\Big(
	\frac{355}{54}
	\!-\!\frac{\pi^2}{18}
	\Big)
      \!-\!\epsilon^2\Big(
	\frac{6523}{324}
	\!-\!\frac{19}{108}\pi^2
	\!-\!\frac{50}{9}\,\zeta(3)
	\Big)
      \Big]
    \bigg\}
    +\mathcal{O}(\epsilon^3)\,,
    \phantom{\Big|}
\\[.4cm]
\bar{F}_{g,\G}^{(1)}&=
  \Big(\frac{\alphas}{4\pi}\Big)\,C_A\,\Big[
    \!-\!\frac{2}{\epsilon^2}
    \!+\!4
    \!+\!\frac{\pi^2}{6}
    \!+\!\epsilon\Big(
      12
      \!+\!\frac{14}{3}\zeta (3)
      \Big)
    \!+\!\epsilon^2\Big(
      28
      \!-\!\frac{\pi^2}{3}
      \!+\!\frac{47}{720}\pi^4
      \Big)
    \Big]
    \!+\!\mathcal{O}(\epsilon^3) \, ,
    \label{eq:gluFFbare1}
\\[.15cm]
\bar{F}_{g,\G}^{(2)}&=
  \Big(\frac{\alphas}{4\pi}\Big)^2 \bigg\{
    C_A^2 \Big[
      \frac{2}{\epsilon^4}
      \!-\!\frac{11}{6\epsilon^3}
      \!-\!\frac{
       \frac{104}{9}
       \!+\!\frac{\pi^2}{6}
       }{\epsilon^2}
      \!-\!\frac{
	\frac{433}{27}
	\!-\!\frac{11}{12}\pi^2
	\!+\!\frac{25}{3}\zeta(3)       
	}{\epsilon}
      \!+\!\frac{3832}{81}
      \!+\!\frac{28\pi^2}{9}
      \!+\!\frac{11}{9}\zeta(3)
      \!-\!\frac{7\pi^4}{60}
      \Big]
\notag \\& \quad
     +C_A N_F \Big[
      \frac{1}{3\,\epsilon^3}
      \!+\!\frac{5}{9\epsilon^2}
      \!-\!\frac{
	\frac{53}{27}
	\!+\!\frac{\pi^2}{6}
	}{\epsilon}
      \!-\!\frac{1591}{81}
      \!-\!\frac{5\pi^2}{18}
      \!-\!\frac{74}{9}\zeta (3)
      \Big]
    +C_F N_F \bigg[
      \!-\!\frac{6}{\epsilon}
      \!-\!\frac{125}{3}
      \!+\!8\zeta(3)
      \Big]
    \bigg\}
    \!+\!\mathcal{O}(\epsilon^1)\,,
    \phantom{\Big|}
\\[.15cm]
\bar{F}_{g,\J}^{\HVBM,(1)}&=
\Big(\frac{\alphas}{4\pi}\Big) \bigg\{
  C_A \Big[
    \!-\!\frac{2}{\epsilon^2}
    \!+\!4
    \!+\!\frac{\pi^2}{6}
    \!+\!\epsilon\Big(
      16
      \!-\!\!\frac{\pi^2}{3}
      \!+\!\frac{32}{3}\zeta(3)
      \Big)
    \!+\!\epsilon^2\Big(
      \frac{152}{3}
      \!-\!\frac{4}{3}\pi^2
      \!+\!2\zeta(3)
      \!+\!\frac{127}{720}\pi^4
      \Big)
    \Big]
\notag\\&\quad
  +C_F \Big[
    2
    \!+\!\epsilon\Big(
      10
      \!-\!12\,\zeta(3)
      \Big)
    +\epsilon^2\Big(
      38
      \!-\!\frac{7}{6}\pi^2
      \!-\!18\,\zeta(3)
      \!-\!\frac{\pi^4}{5}
      \Big)
    \Big]
 \bigg\}
 +\mathcal{O}(\epsilon^3)\, .
\end{align}
\end{subequations}
\normalsize

\subsection{UV renormalization}
\label{sec:UVrenMSbar}

The UV renormalization of the couplings $\alphas$ and $\alphae$ is given
by~\cite{Gnendiger:2014nxa}
\begin{subequations}
\begin{align}
 \bar{Z}_{\alphas}&=1
  +\Big(\frac{\alphas}{4\pi}\Big)\Big\{
    -\frac{\bar{\beta}^{s}_{20}}{\epsilon}\Big\}
  +\Big(\frac{\alphas}{4\pi}\Big)^2
    \Big\{
      \frac{(\bar{\beta}^{s}_{20})^2}{\epsilon^2}
      -\frac{\bar{\beta}^{s}_{30}+\bar{\beta}^{s}_{21}}{2\,\epsilon}
      \Big\}
      +\mathcal{O}(\alphas^3) \, ,
  \label{eq:Zalphas}
  \\
  \bar{Z}_{\alphae}&=1
  +\Big(\frac{\alphas}{4\pi}\Big)
    \Big\{-\frac{\bar{\beta}^{e}_{11}+\bar{\beta}^{e}_{02}}{\epsilon}\Big\}
  +\mathcal{O}(\alphas^2) \, ,
  \label{eq:Zasae}
\end{align}
including the $\beta$ coefficients
\begin{align}
 \bar{\beta}^{s}_{20}&=
    C_A\,\Big(
      \frac{11}{3}
      \!-\!\frac{\epsilon}{3}
      \Big)
    \!-\!\frac{2}{3}N_F\,,
\qquad
    \bar{\beta}^{e}_{11}=
    6\,C_F\,,
\qquad
 \bar{\beta}^{e}_{02}=
    C_A\,\big(
      2
      \!-\!2\epsilon
      \big)
    \!-\!C_F\,\big(
      4
      \!-\!2\epsilon\big)
    -N_F\,,\phantom{|}
\notag\\*
 \bar{\beta}^{s}_{30}&=
    C_A^2\,\Big(
      \frac{34}{3}
      -\frac{14}{3}\epsilon
      \Big)
    -C_A N_F\,\Big(\frac{10}{3}\Big)
    -2\,C_F N_F\,,
    \qquad
 \bar{\beta}^{s}_{21}=
   C_F N_F\,(2\,\epsilon)\,.
\end{align}
\end{subequations}
In Eq.\,\eqref{eq:Zasae}, the renormalized couplings are set equal,
i.\,e.\ $\alphae\!=\!\alphas$. For the calculations in the off-shell case,
also a UV renormalization of the external quark and gluons fields and the gauge
parameter is needed. The corresponding renormalization constants can be found in
Refs.\,\cite{Broggio:2015ata} and \cite{Gnendiger:2016cpg}.

According to operator renormalization in \HVBM, the first perturbative
coefficients of the UV-renormalized form factors in the \FDH\ scheme are
given by
\begin{subequations}
\label{eq:FFren}
\begin{align}
 \bar{\mathcal{F}}_{q,\J}^{\HVBM}
 &=
  \big(1\!+\!\delta\bar{Z}_{\MS}^{\HVBM,(1)}
    \!+\!\delta\bar{Z}_{\MS}^{\HVBM,(2)}\big)
  \big(1\!+\!\delta\bar{Z}_{5}^{\HVBM,(1)}
    \!+\!\delta\bar{Z}_{5}^{\HVBM,(2)}\big)
\notag\\ &\qquad\times
  \big(1\!+\!\bar{F}_{q,\J}^{\HVBM,(1)}
    \!+\!\bar{F}_{q,\J}^{\HVBM,(2)}\big)_{\text{ren.}}
  \!+\!\mathcal{O}(\alphas^3)\,,
  \phantom{\Big|}
  \label{eq:FFren1}
\\[.15cm]
  \bar{\mathcal{F}}_{q,\G}
 &=\frac{
    \big(1+\delta\bar{Z}_{\G\G}^{(1)}\big)
    \big(R_{q,\G}^{(1)}\!+\!\bar{R}_{q,\G}^{(2)}\big)_{\text{ren.}}
      \!+\!\big(\delta\bar{Z}_{\G\J}^{(1)}+\delta\bar{Z}_{\G\J}^{(2)}
      \!\big)
    \big(\bar{R}_{q,\J}^{(0)}\!+\!\bar{R}_{q,\J}^{(1)}\big)
    }{
   R_{q,\G}^{(1)}
   +\,\delta\bar{Z}_{\G\J}^{(1)} R_{q,\J}^{(0)}
   }
  +\mathcal{O}(\alphas^2)\, .
  \label{eq:FFren2}
\\[.15cm]
 \bar{\mathcal{F}}_{g,\G}
 &=\big(1\!+\!\delta\bar{Z}_{\G\G}^{(1)}\!+\!\delta\bar{Z}_{\G\G}^{(2)}\big)
   \big(1\!+\!\bar{F}_{g,\G}^{(1)}\!+\!\bar{F}_{g,\G}^{(2)}\big)_{\text{ren.}}
   \!+\delta\bar{Z}_{\G\J}^{(1)}\,
   \big(R^{(1)}_{g,\J}/R^{(0)}_{g,\G}\big)
  \!+\mathcal{O}(\alphas^3) \, ,
  \phantom{\Big|}
  \label{eq:FFren3}
  \\[.15cm]
 \bar{\mathcal{F}}_{g,\J}^{\HVBM}
 &=\big(1\!+\!\left.\delta\bar{Z}_{5}^{\HVBM}\right.^{(1)}\big)
  \big(1\!+\!\left.\bar{F}_{g,\J}^{\HVBM}\right.^{(1)}\big)_{\text{ren.}}
  \!+\mathcal{O}(\alphas^2)\, ,
  \phantom{\bigg|}
  \label{eq:FFren4}
\end{align}
\end{subequations}
The subscript 'ren.' indicates that the coupling renormalization~%
\eqref{eq:couplingRen} is applied to the bare one-loop amplitudes.
\textit{After} UV renormalization, the evanescent coupling $\alphae$ is
identified with the gauge coupling, i.\,e.\ $\alphae=\alphas$.

\subsection{IR divergence structure}
\label{sec:ir}

The IR divergence structure of one- and two-loop \FDH\ amplitudes has been
investigated in Ref.~\cite{Gnendiger:2014nxa}. Specifying to the case of
massless form factors with two external quarks and gluons, respectively,
a $\mathbf{Z}$ factor subtracting all IR divergences is given by
\begin{subequations}
\label{eq:IRfac}
\begin{align}
      \text{ln}\,\mathbf{Z}_{}&=
      \left(\frac{\alpha_s}{4\pi}\right)\left(
	\frac{\bar{\Gamma}'_{10}}{4\,\epsilon^2}
	+\frac{\bar{\Gamma}_{10}}{2\,\epsilon}\right)
      +\left(\frac{\alpha_e}{4\pi}\right)\left(
	\frac{\bar{\Gamma}'_{01}}{4\,\epsilon^2}
	+\frac{\bar{\Gamma}_{01}}{2\,\epsilon}\right)
      \notag\\&\quad
      +\left(\frac{\alpha_s}{4\pi}\right)^2
      \left(
	-\frac{3\,\betaMS{20}\,\bar{\Gamma}'_{10}}{16\,\epsilon^3}
	+\frac{\bar{\Gamma}'_{20}
	-4\,\betaMS{20}\,\bar{\Gamma}_{10}}{16\,\epsilon^2}
	+\frac{\bar{\Gamma}_{20}}{4\,\epsilon}\right)
      \notag\\
      &\quad\,      
      +\left(\frac{\alpha_s}{4\pi}\right)\left(\frac{\alpha_e}{4\pi}\right)\left(
	-\frac{3\,\betaeMS{11}\,\bar{\Gamma}'_{01}}{16\,\epsilon^3}
	+\frac{\bar{\Gamma}'_{11}-4\,\betaeMS{11}\,\bar{\Gamma}_{01}}{16\,\epsilon^2}
	+\frac{\bar{\Gamma}_{11}}{4\,\epsilon}\right)
      \notag\\
      &\quad\,
      +\left(\frac{\alpha_e}{4\pi}\right)^2\left(
	-\frac{3\,\betaeMS{02}\,\bar{\Gamma}'_{01}}{16\,\epsilon^3}
	+\frac{\bar{\Gamma}'_{02}-4\,\betaeMS{02}\,\bar{\Gamma}_{01}}{16\,\epsilon^2}
	+\frac{\bar{\Gamma}_{02}}{4\,\epsilon}\right)\, .
	\label{eq:IRprediction}
      \end{align}
The relation between the perturbative coefficients of $\text{ln}\,\mathbf{Z}_{}$
and the UV-renormalized form factors is given by
\begin{align}
 \big(\text{ln}\,\mathbf{Z}_{}\big)^{(1)}
 =\bar{\mathcal{F}}_{a,\A}^{\HVBM,(1)}\Big|_{\text{poles}}\,,
 \quad\qquad
  \big(\text{ln}\,\mathbf{Z}_{}\big)^{(2)}
 =\bar{\mathcal{F}}_{a,\A}^{\HVBM,(2)}\Big|_{\text{poles}}
  -\frac{1}{2}\big(\mathcal{F}_{a,\A}^{\HVBM,(1)}\big)^2\Big|_{\text{poles}}\,.
  \label{eq:IRpredictionPert}
\end{align}
\end{subequations}
The $\mathbf{Z}$ factor is written in terms of the IR anomalous dimensions
$\bar{\Gamma}_{mn}'\!=\! -2\,\bar{\gamma}_{mn}^{\text{cusp}}\,C_{q/g}$ and
$\bar{\Gamma}_{mn}\!=\! 2\,\bar{\gamma}_{mn}^{q/g}$ with $C_{q}\!=\!C_F$ for
the quark form factor and $C_{g}=C_A$ for the gluon form factor. 
In \FDH, the values of the partonic IR anomalous dimensions
$\bar{\gamma}_{mn}^{\text{cusp}}$, $\bar{\gamma}_{mn}^{q}$, and
$\bar{\gamma}_{mn}^{g}$ are known up to the two-loop level~\cite{Gnendiger:2014nxa}.
Together with the known values of the one-loop $\beta$ coefficients it is
therefore possible to predict the entire IR divergence structure of the \FDH\
form factors up to the two-loop level. Since Eq.\,\eqref{eq:IRprediction} is
written in terms of UV renormalized couplings, they can be set equal
($\alphae=\alphas$).

\bigskip
\section*{Acknowledgments} 
It is a pleasure to thank Dominik St$\hat{\text{o}}$ckinger and
Michael Spira for fruitful discussions and helpful comments on the manuscript.
\bigskip

\bibliography{bibliography}{}
\bibliographystyle{JHEP}

\end{document}